\documentclass[twocolumn]{aastex701}
\usepackage{CJKutf8}
\usepackage{threeparttable}
\usepackage{multirow}
\definecolor{mygreen}{RGB}{1,1,1}

\newcommand{\tablenotetextinline}[2]{%
  {\currtabletypesize
   \hskip1em 
   $^{\hbox to 5pt{$#1$}}$#2}%
}

\begin{document}

\title{Onset of CN Emission in 3I/ATLAS: Evidence for Strong Carbon-Chain Depletion}

\author[0000-0002-6514-318X]{Luis E. Salazar Manzano}
\affiliation{Department of Astronomy, University of Michigan, Ann Arbor, MI 48109, USA}
\email[show]{lesamz@umich.edu}  

\author[0000-0001-7737-6784]{Hsing~Wen~Lin(\begin{CJK*}{UTF8}{bkai}林省文\end{CJK*})}
\affiliation{Department of Physics, University of Michigan, Ann Arbor, MI 48109, USA}
\affiliation{Michigan Institute for Data and AI in Society, University of Michigan, Ann Arbor, MI 48109, USA}
\email{hsingwel@umich.edu}

\author[0000-0002-0140-4475]{Aster G. Taylor}
\altaffiliation{Fannie and John Hertz Foundation Fellow}
\affiliation{Department of Astronomy, University of Michigan, Ann Arbor, MI 48109, USA}
\email{agtaylor@umich.edu}

\author[orcid=0000-0002-0726-6480]{Darryl Z. Seligman}
\altaffiliation{NSF Astronomy and Astrophysics Postdoctoral Fellow}
\affiliation{Department of Physics and Astronomy, Michigan State University, East Lansing, MI 48824, USA}
\email{dzs@msu.edu} 

\author[0000-0002-8167-1767]{Fred C. Adams}
\affiliation{Department of Physics, University of Michigan, Ann Arbor, MI 48109, USA}
\affiliation{Department of Astronomy, University of Michigan, Ann Arbor, MI 48109, USA}
\email{fca@umich.edu}

\author[0000-0001-6942-2736]{David W. Gerdes}
\affiliation{Department of Physics, Case Western Reserve University, Cleveland, OH 44106, USA}
\affiliation{Department of Physics, University of Michigan, Ann Arbor, MI 48109, USA}
\affiliation{Department of Astronomy, University of Michigan, Ann Arbor, MI 48109, USA}
\email{gerdes@umich.edu}

\author[orcid=0000-0003-0403-0891]{Thomas Ruch}
\affiliation{Department of Physics, University of Michigan, Ann Arbor, MI 48109, USA}
\email{trruch@umich.edu}

\author[0009-0000-4697-5450]{Tessa T. Frincke}
\affiliation{Department of Physics and Astronomy, Michigan State University, East Lansing, MI 48824, USA}
\email{frincket@msu.edu}

\author[0000-0003-4827-5049]{Kevin J. Napier} 
\affiliation{Center for Astrophysics $\mid$ Harvard \& Smithsonian, Cambridge, MA 02138, USA}
\affiliation{Michigan Institute for Data and AI in Society, University of Michigan, Ann Arbor, MI 48109, USA}
\affiliation{Department of Physics, University of Michigan, Ann Arbor, MI 48109, USA}
\email{kevin.napier@cfa.harvard.edu}

\begin{abstract}

Interstellar objects provide a direct window into the environmental conditions around stars other than the Sun. The recent discovery of 3I/ATLAS, \textcolor{mygreen}{a new interstellar comet}, offers a unique opportunity to investigate the physical and chemical properties of interstellar objects and to compare them with those of comets in our own Solar System. In this Letter we present the results of a 10-night spectroscopic and photometric monitoring campaign with the 2.4 m Hiltner and 1.3 m McGraw-Hill telescopes at the MDM Observatory. The campaign was conducted between August 8 and 17 while 3I/ATLAS was inbound at heliocentric distances of 3.2 - 2.9 au. Our observations captured the onset of \textcolor{mygreen}{optical gas activity}. Nightly spectra reveal a weak CN emission feature in the coma of 3I/ATLAS, absent during the first nights but steadily strengthening thereafter. We measure a CN production rate of \textcolor{mygreen}{$Q$(CN)$\sim6\times$10$^{24}$ s$^{-1}$}, towards the lower end of activity observed in Solar System comets. Simultaneous photometry also indicates a small but measurable increase in the coma’s radial profile and increasing $r$-band $Af\rho$ with values in the order of $\sim300$ cm. We derived a gas-to-dust production ratio of $\log Q (\mathrm{CN})/Af\rho\sim\textcolor{mygreen}{22.4}$. Our upper limit on the C$_2$-to-CN ratio ($\log Q(\mathrm{C}_2)/Q(\mathrm{CN})\lesssim\textcolor{mygreen}{-0.8}$) indicates that 3I/ATLAS is a strongly carbon-chain depleted comet.  Further observations of 3I/ATLAS are required to verify the apparent carbon-chain depletion and to explore whether such composition represents a recurring trait of the interstellar comet population.
\end{abstract}



\section{Introduction} \label{sec:intro}

Comets are composed of a mixture of ices, including H$_2$O, CO, and CO$_2$ \citep{2015SSRv..197....9C, 1950ApJ...111..375W}, together with dust in the form of silicates (amorphous and crystalline) and refractory organics \citep{2018SSRv..214...64L}. While most of our understanding of comets comes from objects within our Solar System, the discovery of interstellar comets has opened a new window for their study \citep{Fitzsimmons2024}. The volatile component is the primary driver of cometary activity \citep{2004come.book..523C}, and the molecular composition of the outgassed material provides a powerful diagnostic on both the formation environment and the evolutionary processing of comets  \citep{2011ARA&A..49..471M, 2015SSRv..197..191D, 2015SSRv..197..271G, 2017RSPTA.37560252B}. For example, carbon-to-oxygen ratios can serve as tracers of formation locations within protoplanetary disks \citep{2022PSJ.....3..150S}. Beyond the dust continuum produced by scattered sunlight, the dominant features in optical cometary spectra arise from radicals such as CN, C$_3$, and C$_2$ \citep{2004come.book..425F, 2000Icar..147..180F}. The fluorescence emission of these molecules originates from the photodissociation of parent species released from the nucleus.

The CN violet system $B\,{}^{2}\Sigma^{+} - X\,{}^{2}\Sigma^{+}$, centered near $\sim 3870~\text{\AA}$, is one of the most prominent features in optical cometary spectra \citep{2004come.book..425F}. As the daughter product of HCN photodissociation \citep{2005P&SS...53.1243F, 2011ApJ...741...89I}, CN is typically detected in comets inside a few au, where overall activity (often water-driven) rises, though it can also appear at larger heliocentric distances when more volatile species (e.g., CO or CO$_2$) drive the outgassing. This feature has been extensively used in surveys of Solar System comets, which have revealed the existence of compositional classes, originally categorized as ``typical'' and ``carbon-chain depleted'' \citep{1995Icar..118..223A}, with evidence for further  subclasses \citep{2012Icar..218..144C, 2014acm..conf..475S}. Carbon chain-depleted comets are found across all dynamical classes, but they make up a larger fraction of Jupiter Familiy Comets (JFC) compared to Long Period Comets (LPCs). Whether these differences arise from evolutionary processing or distinct formation environments remains uncertain, although studies of \textcolor{mygreen}{the fragmented} comet 73P/Schwassmann-Wachmann 3 strongly suggest they reflect primordial composition \citep{2011AJ....141..177S}.  


The first two interstellar objects were 1I/`Oumuamua and 2I/Borisov, discovered in 2017 \citep{Williams17} and 2019 \citep{borisov_2I_cbet} respectively. There were a set of upper limits of gas and dust production rates \citep{Meech2017,Jewitt2017} measured for 1I/`Oumuamua, which was photometrically inactive. These upper limits, summarized in Table 3-4 of \citet{Jewitt2023ARAA}, were placed on CN, C$_2$, C$_3$ \citep{Ye2017}, OH \citep{Park2018}, CO, and CO$_2$ \citep{Trilling2018}. These nondetections of spectral emission features and dust were coupled with a significant detection of comet-like nongravitational acceleration in the radial direction \citep{Micheli2018}. This combination led to a variety of hypotheses regarding the provenance of 1I/`Oumuamua \citep{Micheli2018,Sekanina2019,MoroMartin2019,Flekkoy19,Luu20,Seligman2020,Levine2021,Levine2021_h2,desch20211i,jackson20211i,Desch2022,Bergner2023}. Although strong nongravitational accelerations have been since identified on a set of 14 near-Earth objects that also do not display dust activity \citep{Farnocchia2023,Seligman2023b,Seligman2024PNAS}, the nature of 1I/`Oumuamua is still uncertain.


As opposed to the uncertainity in 1I/`Oumuamua's origin, the apparition of 2I/Borisov was amenable to observations from ground and space based facilities, which provided detailed composition data throughout the apparition. A suite of spectral observations were obtained that characterized the volatile outgassing composition pre- and post-perihelion. For a comprehensive list see Table 4 and 5 of \citet{Jewitt2023ARAA}. In particular, H$_2$O, CO, OH, \textcolor{mygreen}{HCN,} CN, C$_2$ and C$_3$  production rates were measured \citep{Xing2020,Bodewits2020,Cordiner2020,Opitom2019,Kareta2019,McKay2020,Lin2020,Bannister2020,Aravind2021,yang2021}. For reviews on the topic of interstellar objects see \citet{Fitzsimmons2024}, \citet{Seligman2023}, \citet{Jewitt2023ARAA}, and \citet{MoroMartin2022}.

3I/ATLAS, the third interstellar object discovered, is an active comet. A large number of follow-up observations have been conducted after the discovery \citep{Denneau2025} by the Asteroid Terrestrial-impact Last Alert System (ATLAS) survey \citep{Tonry2018a}. The object was immediately reported to display activity \citep{2025ATel17263....1J,2025ATel17264....1A,Seligman2025} including in precovery observations \citep{Chandler2025,Feinstein2025,Martinez-Palomera2025}. Based on the kinematics, 3I/ATLAS is 3--11 Gyr old and therefore provides a tracer of stellar system formation in the early Galaxy \citep{2025arXiv250705318H, Taylor2025a}. The nucleus has an estimated diameter between 0.22 and 2.8 km \citep{2025arXiv250802934J}. Photometric light curve measurements indicate rotational period of $16.79 \pm 0.23$ hours \citep{2025A&A...700L...9D,2025arXiv250800808S}. The optical colors are broadly similar to D-type asteroids \citep{Opitom2025,2025RNAAS...9..194B,Kareta2025,Beniyama2025}. The near-infrared spectrum \textcolor{mygreen}{shows} clear evidence of water ice, \textcolor{mygreen}{H$_2$O, CO and CO$_2$} in the coma \citep{Yang2025,Lisse2025,Cordiner2025}, while UV observations reveal the presence of OH and Ni \citep{Xing2025, Rahatgaonkar2025}. Although the presence of volatiles such as H$_2$O and CO$_2$ can account for its active behavior, no optical gas emission had been detected prior to August 2025. 

Assessing gas emission is essential for understanding the formation environment and history of 3I/ATLAS. Because gas release is time-dependent, fluorescence emission was expected to become detectable in the optical as the comet approached 3 au from the Sun in August 2025, provided the parent molecules were present. Motivated by this, we carried out spectroscopic and photometric monitoring of 3I/ATLAS between August 8 and 17 to search for signs of optical activity. In this Letter we report the onset of CN emission in its optical spectrum, which emerged in mid-August as independently confirmed by the VLT \citep{Rahatgaonkar2025} and the Lowell Observatory \citep{2025ATel17352}. Section~\ref{sec:observations} describes our spectroscopic and photometric monitoring conducted with the MDM Observatory. Section~\ref{sec:reduction} outlines the processing algorithms for both data sets. Section~\ref{sec:results} presents our CN detection, its temporal evolution, constraints on molecular production rates, and the accompanying photometric behavior. Section~\ref{sec:discussion} considers the implications for the compositional classification of 3I/ATLAS. Finally, Section~\ref{sec:conclusion} summarizes our main results.

\section{Observations} \label{sec:observations}

\begin{deluxetable*}{cccc|cc|ccc}
\tablecolumns{9}
\tablecaption{Log of Observations.}
\label{tab:obs}
\tablewidth{0pt}
\tablehead{
\multirow{2}{*}{Night (UTC-7)} & 
\multirow{2}{*}{$r_h$ [au]\tablenotemark{a}} & 
\multirow{2}{*}{\textcolor{mygreen}{$\dot{r}$ [km\ s$^{-1}$]}\tablenotemark{b}} & 
\multirow{2}{*}{$\Delta$ [au]\tablenotemark{c}} &
\multicolumn{2}{c|}{2.4 m telescope} &
\multicolumn{3}{c}{1.3 m telescope} \\
& & & & Exp Time [s] & Airmass & Exp Time [s] & Airmass & Bands
}
\startdata
2025 Aug 09 & 3.19 & \textcolor{mygreen}{-55.46} & 2.70 & $4 \times 120$, $11 \times 300$ & 1.57--2.04 & $38 \times 60$ & 1.69--2.09 & r \\
2025 Aug 10 & 3.16 & \textcolor{mygreen}{-55.35} & 2.69 & $11 \times 300$ & 1.67--2.08 & -- & -- & -- \\
2025 Aug 12 & 3.10 & \textcolor{mygreen}{-55.13} & 2.67 & $15 \times 300$ & 1.57--2.15 & $52 \times 60$ & 1.54--2.09 & r \\
2025 Aug 16 & 2.97 & \textcolor{mygreen}{-54.65} & 2.65 & $9 \times 300$ & 1.79--2.15 & -- & -- & -- \\
2025 Aug 17 & 2.94 & \textcolor{mygreen}{-54.51} & 2.64 & $6 \times 300$ & 1.67--1.82 & 3, 9, $5 \times 120$ & 1.57--1.88 & U, g, r \\
\enddata
\vspace{1pt}
\tablenotetextinline{a}{Heliocentric distance.}\tablenotetextinline{b}{\textcolor{mygreen}{Radial heliocentric velocity.}}\tablenotetextinline{c}{Geocentric distance.}
\end{deluxetable*}

Observations were conducted at the MDM Observatory on Kitt Peak, which operates the 2.4\,m Hiltner telescope and the 1.3\,m McGraw-Hill telescope. The 2.4\,m was used for spectroscopic characterization, while the 1.3\,m provided simultaneous photometric monitoring.

\subsection{Instrumental setups} \label{subsec:telescopes}

The 2.4\,m observations were obtained with the Ohio State Multi-Object Spectrograph (OSMOS; \citealt{2011PASP..123..187M, 2010SPIE.7735E..4LS}). OSMOS was used with the MDM4K CCD detector, which has a plate scale of 0\farcs273 per pixel. To reduce the readout time (to $\sim$30\,s), only a 4k$\times$1k subsection of the chip was read out, providing a field of view of $\sim$17\arcmin. The CCD quantum efficiency peaks at 90\% near 600\,nm and remains above 60\% across 300–850\,nm.

For the OSMOS spectroscopy we employed the triple-prism configuration, which provides broad wavelength coverage ($\sim$3500-10,000\,\AA) at the expense of low resolution. In this setup the spectral resolution \textcolor{mygreen}{varies strongly with wavelength (see Figure 4 from \citealt{2011PASP..123..187M}), from $\lambda/\Delta\lambda \sim 400$ in the blue ($\Delta\lambda\sim10$ \AA\ at 4000 \AA) to $\sim 60$ in the red ($\Delta\lambda\sim100$ \AA\ at 8000 \AA)}. Because of the limited resolution, internal arc lamps were not suitable for wavelength calibration; instead, we calibrated the dispersion solution using the spectrum of a well-characterized planetary nebula. All spectroscopic observations were obtained \textcolor{mygreen}{in the long-slit mode, using a 1\farcs4-wide slit}.

Simultaneous photometry with the 1.3\,m telescope was obtained using the Templeton imager, a 1024$\times$1024 CCD camera. Observations were carried out with SDSS $g$ and $r$ filters and the standard Johnson $U$ filter.

\subsection{Observing run} \label{subsec:strategy}

The MDM observing campaign reported in this Letter took place from 2025 August 8 through August 17, spanning 10 nights in total (Table~\ref{tab:obs}). As 3I/ATLAS approached perihelion, the target was observable only in a limited window between the end of astronomical twilight and the point where its airmass exceeded 2. Early in the run, this window lasted about 1.5\,h but was affected by contamination from the full Moon, whereas during the final nights the window shortened to $\sim$1\,h but was free of lunar interference for most of the science exposures and calibration frames.

The 2.4\,m telescope does not support non-sidereal tracking. Given the relative high rate of motion ($100\arcsec\ \mathrm{hr}^{-1}$) and to minimize recentering overheads, we rotated OSMOS so that the slit was aligned with the trajectory of 3I/ATLAS. The required rotation angle, $\sim$80$^\circ$ east, was computed from the position angle provided by the MPC. To place the target on the slit, we first obtained unfiltered acquisition images of the field and then used MDM's \texttt{osctrtask.py} script to determine the offsets for the guiding camera. With this procedure, centering accuracy was typically better than $\pm$1 pixel. \textcolor{mygreen}{During a typical 5-minute exposure, the comet drifted by $\sim8\arcsec$ along the slit, which is only a small fraction of the $5\arcmin$ effective image height}.

The target acquisition and observing sequences varied slightly from night to night depending on atmospheric conditions. Each night began with twilight flats, followed by acquisition images of the target field to confirm the pointing. On some nights, field recognition and slit centering were completed before the end of astronomical twilight, and 3I/ATLAS spectra started to be obtained despite the residual sky brightness in order to maximize the number of science exposures. After the science frames, we observed calibration standards chosen from a catalog in the same sky region and at similar airmass. Flux standards were drawn from the spectrophotometric catalog of \citet{1990AJ.....99.1621O}. For wavelength calibration we used compact planetary nebulae ($<$10\arcsec\ in angular size) selected from \citet{1992secg.book.....A} and \citet{2001A&A...378..843K}. Solar analogs were selected from the Hipparcos catalog, restricting to G2V stars of sufficiently low brightness to avoid CCD saturation.

Observing conditions and instrumental performance varied considerably over the 10-night campaign. We obtained data on Saturday 9, Tuesday 12, and Sunday 17 using the 1.3\,m imager. $U$, $g$, and $r$ photometry were acquired only on the 17th, since we took only $r$-band images on the other two nights. For the 2.4\,m spectroscopy, we focus on the highest-quality datasets obtained on Sunday 10, Tuesday 12, Saturday 16, and Sunday 17. Conditions on the 10th and 12th were acceptable but not photometric, while the nights of the 16th and 17th were photometric. On the 17th, however, observations were interrupted by a passing cloud midway through the window, preventing further exposures of 3I/ATLAS as well as calibration frames. For this reason, we applied the calibration data obtained on the 16th to the spectra from the 17th, justified by the similar sky conditions on both nights.

\section{Data processing} \label{sec:reduction}

The spectroscopic and photometric data were reduced using custom Python pipelines, which we describe below. The spectra processing routines are based on the algorithms presented in \citet{Lin2020}.

\subsection{Spectroscopy}\label{subsec:spec}

For the 2.4\,m spectroscopy, the calibration sequence included bias subtraction, wavelength calibration, extinction correction, flux calibration, and division by a solar-analog spectrum. Bias subtraction was performed with MDM's \texttt{proc4k.py} script, which estimates the bias level from the overscan region and subtracts it from each of the four amplifier regions of the CCD. \textcolor{mygreen}{For wavelength calibration, we identified ten strong and spectrally resolved atomic and ionized emission lines (including all of the hydrogen Balmer lines) from the 1D spectrum of the planetary nebula. The line peak wavelengths, measured as a function of CCD pixel position, were fit with a low-order polynomial to derive the wavelength solution, producing residuals of $<1$\,nm}. \textcolor{mygreen}{Atmospheric extinction of the 1D spectra for the solar analogues, flux standards, and the science target was corrected using the mean vertical extinction curve for Kitt Peak}, with airmass approximated as the secant of the zenith distance. The response function was then determined from the extinction-corrected spectrum of the flux standard and applied to the extinction-corrected spectra of 3I/ATLAS. Finally, the reflectance spectrum of 3I/ATLAS was obtained by dividing the flux-calibrated 1D spectrum by that of the solar analog.

To extract the 1D spectra from the 2D frames, we first identified the source trace by fitting a Gaussian to the spatial profile of the spectrum collapsed along the wavelength direction. This fit also provided the FWHM, which we used to define the sizes of three apertures: the source aperture, the sky apertures on either side of the trace, and the gap separating the source and sky regions. Typical extractions \textcolor{mygreen}{of the calibration sources (planetary nebula, flux standards, and solar analogs) adopted source apertures of $\sim$1–2 FWHM, sky apertures of $\sim$0.5–1.5 FWHM}, and a gap of $\sim$0.5 FWHM; larger gaps were used in cases where extended wings of the spatial profile were not well represented by the Gaussian fit. For the extraction of 3I/ATLAS, we fixed the source aperture radius to correspond to a projected distance of 15000 km at the comet for each day. Flux in the source aperture was averaged, while the flux in the sky apertures was median-combined, and the two sky values were subsequently averaged if both were used. Each extracted 1D spectrum was then passed through a cosmic-ray rejection routine, in which points deviating more than 10$\sigma$ above the local continuum were clipped. Multiple 1D spectra of the same object obtained on a given night, after extinction and flux calibration, were co-added using a weighted average with weights proportional to the total signal of each spectrum.

\subsection{Imaging/Photometry}\label{subsec:phot}

The 1.3\,m images were reduced using standard CCD procedures, including bias subtraction and flat-field calibration. Photometric zero-points were determined for each frame by calibrating against background stars matched to the SkyMapper catalog. Because the filter systems differ between SDSS and SkyMapper, we applied the transformations from \citet{skymapper_dr1} to convert SkyMapper $g_{\rm sm}$ and $r_{\rm sm}$ magnitudes to the SDSS $g$ and $r$ system. For the Johnson $U$ band, no direct transformation exists with the closest SkyMapper filter, $u_{\rm sm}$. However, since their effective wavelengths ($U$: 3650\,\AA; $u_{\rm sm}$: 3500\,\AA) and bandpasses ($U$: 3200–4000\,\AA; $u_{\rm sm}$: 3000–3800\,\AA) are similar \citep{skymapper_filters}, we compare the $U$ filter to $u_{\rm sm}$ to calculate the zero-point of the $U$-band images.

\section{Results} \label{sec:results}

\subsection{3I/ATLAS reflectance spectrum}\label{subsec:reflectance}

\begin{figure}[h]
    \centering
    \includegraphics[width=1\linewidth]{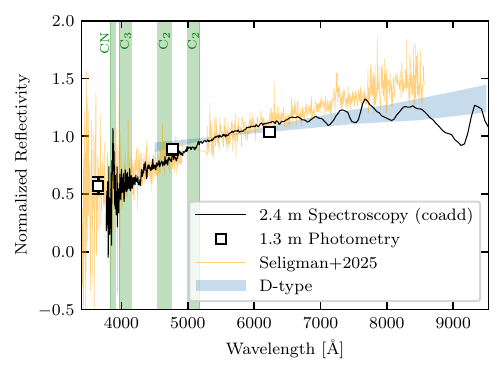}
    \caption{Mean reflectance spectrum of 3I/ATLAS, co-added from the nights of August 10, 12, 16, and 17 with the 2.4\,m MDM telescope (black). For comparison, we show the spectra of 3I/ATLAS reported by \citet{Seligman2025} (yellow), obtained on July 4 with the 2.2\,m UH telescope, and the D-type spectral taxonomy from \citet{2009Icar..202..160D} (blue). Squares mark photometric measurements obtained with the 1.3\,m MDM telescope. Vertical bands mark the positions of typical cometary emission features: CN at 3870 $\mathrm{\AA}$, C$_3$ at 4050 $\mathrm{\AA}$, C$_2$($\Delta\nu=1$) at 4750 $\mathrm{\AA}$, and  C$_2$($\Delta\nu=0$) at 5150 $\mathrm{\AA}$.}
    \label{fig:reflectance}
\end{figure}

A reflectance spectrum was computed independently for each of the four spectroscopic nights (August 10, 12, 16, and 17), and then averaged to produce the mean spectrum shown in Figure~\ref{fig:reflectance}. \textcolor{mygreen}{For the coadded spectrum all nights were weighted equally, except August 10, which was down-weighted by a factor of 1/10 due to its higher background noise contamination}. The short-wavelength slope of our co-added spectrum is consistent with values reported by other facilities. Between 3900–6000\,\AA\ we measure a slope of 26.1$\pm$0.5\%/k\AA, in agreement with the 27.4$\pm$1.0\%/k\AA\ reported by \citet{2025ApJ...990L..27P} over 4000-5500 \,\AA. At longer wavelengths (5500–7000\,\AA) we derive a shallower slope of 14.0$\pm1.9$\%/k\AA, consistent with a D-type spectral taxonomy \citep{2009Icar..202..160D}. The red portion of our spectrum is slightly shallower than the red spectra part from \citet{Seligman2025} (see Figure~\ref{fig:reflectance}), \citet{2025A&A...700L...9D}, or \citealt{2025ApJ...990L..27P}, the latter of whom measured a slope of 16.4$\pm$0.4\%/k\AA\ in the 5500–7000\,\AA\ range. However, our spectra \textcolor{mygreen}{between 7000–8500\,\AA\ has a slope of 7.4$\pm$2.2\%/k\AA\ consistent} with the value of 5.2$\pm$0.2\%/k\AA\ reported by \citet{Yang2025} in the 7000–9000\,\AA\ interval with GMOS.

\subsection{CN emission}\label{subsec:CN}

\begin{figure*}[ht]
    \centering
    \includegraphics[width=1\linewidth]{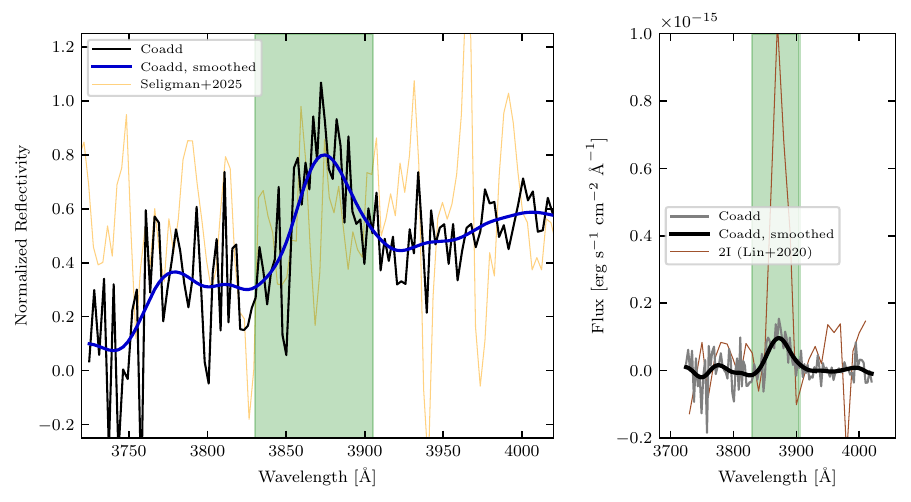}
    \includegraphics[width=1\linewidth]{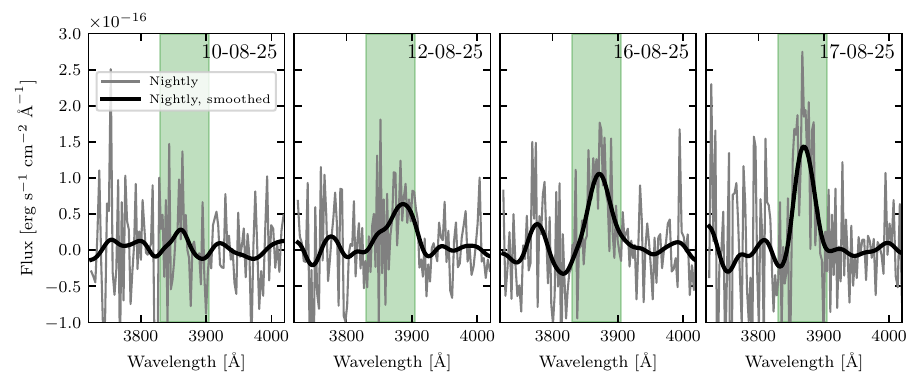}
    \caption{CN emission feature of 3I/ATLAS detected with the 2.4~m MDM telescope. The vertical green band mark the expected position of CN emission near 3870~$\mathrm{\AA}$. \textbf{Top left:} Zoom-in on the CN region of the reflectance spectrum. In addition to the comparison spectra shown in Figure~\ref{fig:reflectance}, we include a Gaussian-smoothed curve to improve visibility (blue). \textbf{Top right:} Co-added, continuum-subtracted flux of 3I/ATLAS from this work (gray), compared with the CN emission flux detected in 2I/Borisov by \citet{Lin2020} (brown). A Gaussian smoothing applied to our data is also shown (black). \textbf{Bottom:} Continuum-subtracted CN flux measured individually on August 10, 12, 16, and 17.}    
    \label{fig:CN}
\end{figure*}

A clear CN emission feature is visible in our mean reflectance spectrum of 3I/ATLAS (Figure~\ref{fig:reflectance}), and a zoom-in of this region is shown in the top-left panel of Figure~\ref{fig:CN}. \textcolor{mygreen}{We also show the spectra convolved with a Gaussian function of $\sigma=4$ pixels.} Although the CN feature was not reported during the extensive spectroscopic monitoring of 3I/ATLAS in July following its discovery \citep{Seligman2025, 2025ApJ...990L..27P, 2025arXiv250800808S, 2025A&A...700L...9D, 2025A&A...700L..10A}, our observations in mid-August with the 2.4\,m MDM telescope indicate that \textcolor{mygreen}{outgassing of CN-bearing parent volatiles is underway, producing detectable levels of CN emission in the coma.}

To quantify the strength of the CN band, we isolated the emission flux from the flux-calibrated 1D spectra. \textcolor{mygreen}{First, for each night we subtracted the scaled and reddened solar analogue spectrum to remove the underlying dust continuum. We then} selected a 300\,\AA\ window centered on 3870\,\AA, masked the CN emission feature, and fitted a low-order polynomial to bring the continuum to a zero-flux level. The resulting continuum-subtracted spectra were subsequently combined to produce the co-added emission spectrum shown in the top right panel of Figure~\ref{fig:CN}, while the nightly extractions are displayed in the lower panels of the same figure.

Our nightly data may suggest an evolution of the CN emission over the course of our week-long monitoring campaign (bottom panels of Figure~\ref{fig:CN}). It is important to note that the measurements from August 10 and 12 were obtained under significantly poorer and less stable sky conditions compared to the observations on the 16th and 17th. Thus, we cannot completely rule out the possibility that the weak CN detections on the first two nights were partly affected by atmospheric conditions. Nevertheless, we clearly observed an increase in CN emission amplitude between the 16th and 17th, particularly since both nights were conducted under photometric conditions and free of scattered moonlight. Our first CN detection on August 12 (August 13 UTC) is one day before to the first CN detection reported \citet{Rahatgaonkar2025} on August 14, and 6 days before the detection by \citet{2025ATel17352} on August 19. \textcolor{mygreen}{A faint feature is present at the expected CN wavelength range in the August 10 data, though the noise level is too high to confidently classify it as a detection.}

The total CN emission on each night with a detection was quantified by fitting \textcolor{mygreen}{a line model} to the continuum-subtracted emission feature. \textcolor{mygreen}{Although the CN violet band is intrinsically asymmetric, we adopted a Gaussian profile since, at the low resolution of the triple-prism setup, the line is unresolved and relatively weak}. For August 12, 16, and 17, we measured fluxes of \textcolor{mygreen}{$(3.04\pm0.22)\times10^{-15}$, $(4.65\pm0.21)\times10^{-15}$, and $(5.47\pm0.29)\times10^{-15}$ erg s$^{-1}$ cm$^{-2}$}, respectively. For August 10 we obtained a 3$\sigma$ upper limit \textcolor{mygreen}{(using the procedure from \citealt{2012Icar..218..144C})} in the CN flux of \textcolor{mygreen}{$<(3.40\pm1.13)\times10^{-15}$} erg s$^{-1}$ cm$^{-2}$. \textcolor{mygreen}{If we consider the August 10 data as a detection, it yields a flux of $(9.1\pm2.1)\times10^{-16}$ erg s$^{-1}$ cm$^{-2}$.}

\textcolor{mygreen}{We converted the CN fluxes within our rectangular aperture into molecular production rates using the standard formulation \citep{2004come.book..425F, 2004come.book..523C}, and to extrapolate the production rate to the entire coma we used the Haser model \citep{1957BSRSL..43..740H, 2020PSJ.....1...83H}. For this calculation, we used CN fluorescence efficiencies from \citealt{2010AJ....140..973S}, scale lengths from \citealt{1995Icar..118..223A}, and assumed an outflow velocity of 1 km s$^{-1}$}. The resulting CN production rates $Q$(CN) \textcolor{mygreen}{(see Table~\ref{tab:physical})} are \textcolor{mygreen}{$(4.82\pm0.34)\times10^{24}$, $(6.34\pm0.29)\times10^{24}$, and $(7.17\pm0.38)\times10^{24}$} s$^{-1}$ for August 12, 16, and 17, respectively. Our upper limit for August 10 is $Q$(CN) \textcolor{mygreen}{$<(5.8\pm1.9)\times10^{24}$} s$^{-1}$\textcolor{mygreen}{, but if this night is treated as a detection, it yields $(1.6\pm0.4)\times10^{24}$ s$^{-1}$}. 

\textcolor{mygreen}{The extrapolation of our $Q$(CN) values is consistent with the estimate from \citet{2025ATel17352} on August 19 (UTC) from the Lowell Observatory. \citet{Rahatgaonkar2025} report an order of magnitude lower values on August 14-21 (UTC) from VLT UVES and X-shooter, however, note that they use a different outflow velocity prescription.} Our detection measurements over the 5-day detection window are consistent with a heliocentric distance dependence following a power-law index of 
\textcolor{mygreen}{$-7.3\pm1.1$ when using the August 12, 16, and 17 detections, consistent within uncertainties with the $-9.38\pm1.20$} measurement from \citet{Rahatgaonkar2025}. \textcolor{mygreen}{This power-law index is relatively steep compared to CN production rate dependencies observed in Solar System comets, whose population averages fall between -0.93 and -4.5 \citep{1995Icar..118..223A}, suggesting a significant increase on CN production for future measurements}. As both \citet{Rahatgaonkar2025} and \citet{2025ATel17352} appeared while this Letter was in the late stages of preparation, a fuller comparison is beyond our scope. 

Our co-added MDM spectra of 3I can be directly compared with the CN emission detected in 2I/Borisov, which was observed with the same instrument configuration on the MDM 2.4 m telescope \citep{Lin2020}. Both objects show CN emission, \textcolor{mygreen}{and although the CN production rate is higher compared to 2I/Borisov,} the co-added line flux of 3I/ATLAS is significantly weaker (top right panel of Figure~\ref{fig:CN}). This difference could reflect the different heliocentric distances at which they were observed. The MDM 2I/Borisov spectra were obtained at $r_h \sim 2.1$ au, whereas our 3I/ATLAS observations were conducted nearly 1 au farther out, at $r_h \sim 3.0$~au.

\begin{deluxetable*}{lccccc}
\tablecolumns{6}
\tablecaption{Derived physical parameters.}
\label{tab:physical}
\tablewidth{0pt}
\tablehead{
Night (UTC-7) & $Q$(CN) [s$^{-1}$] & $Af\rho$ [cm]\tablenotemark{\footnotesize\string a} & $r$ magnitude & $Q(\mathrm{C}_3)$ [s$^{-1}$] & $Q(\mathrm{C}_2)$ [s$^{-1}$]\tablenotemark{\footnotesize\string b}
} 
\startdata
2025 Aug 09 & - & $ 260 \pm 20$ & $ 16.65 \pm 0.06$ & - & -  \\
2025 Aug 10 &  $ \textcolor{mygreen}{<(5.8\pm1.9) \times 10^{24}}$ & - & -  & \textcolor{mygreen}{$<(12.0\pm3.9)\times10^{23}$} & \textcolor{mygreen}{$<(2.47\pm0.82)\times10^{24}$}\\
2025 Aug 12 & $ \textcolor{mygreen}{(4.82\pm0.34) \times 10^{24}}$ & $300 \pm 10$ & $16.42 \pm 0.03$ & \textcolor{mygreen}{$<(8.9\pm3.0)\times10^{23}$} & \textcolor{mygreen}{$<(2.84\pm0.95)\times10^{24}$} \\
2025 Aug 16 & $ \textcolor{mygreen}{(6.34\pm0.29) \times 10^{24}}$ & - & - & \textcolor{mygreen}{$<(8.9\pm3.0)\times10^{23}$} & \textcolor{mygreen}{$<(1.43\pm0.48)\times10^{24}$}\\
2025 Aug 17 & $ \textcolor{mygreen}{(7.17\pm0.38) \times 10^{24}}$ & $315 \pm 2$ & $ 16.24 \pm 0.01$ & \textcolor{mygreen}{$<(9.4\pm3.1)\times10^{23}$} & \textcolor{mygreen}{$<(1.43\pm0.48)\times10^{24}$} \\
\enddata
\vspace{1pt}
\tablenotetextinline{a}{$r$ band dust production rate.}\tablenotetextinline{b}{$\Delta\nu=0$.}
\end{deluxetable*}

\subsection{C$_3$ and C$_2$ constraints}\label{subsec:carbon}

Despite \textcolor{mygreen}{the fact that} C$_3$ and C$_2$ are commonly found in comets \citep{2005A&A...442.1107H, 2012A&A...538A.149W}, we do not have a confident detection of these emission lines in our optical spectrum of 3I/ATLAS (Figure~\ref{fig:reflectance}). We obtained C$_3$ and C$_2(\Delta \nu=0)$ upper limits using a procedure similar to that used to compute CN upper limits in the previous section \textcolor{mygreen}{and adopting the wavelength bandpasses defined by \cite{2011Icar..213..280L}.} For C$_3$ we obtained upper limits of \textcolor{mygreen}{$(6.6\pm2.2)$, $(5.3\pm1.8)$, $(6.1\pm2.1)$ and $(6.7\pm2.2)$ $\times10^{-15}$} erg s$^{-1}$ cm$^{-2}$ for the nights of August 10, 12, 16 and 17, respectively. On the other hand, we obtained  C$_2(\Delta \nu=0)$ flux upper limits of \textcolor{mygreen}{$(1.35\pm0.45)$, $(1.68\pm0.56)$, $(0.99\pm0.33)$, $(1.03\pm0.34)$ $\times10^{-15}$} erg s$^{-1}$ cm$^{-2}$. The corresponding upper limits in the molecular production rates of these species are presented in Table~\ref{tab:physical}. \textcolor{mygreen}{Additionally, we derived upper limits from the coadded spectrum. For C$_3$ and C$_2$, we obtained, respectively, flux upper limits of $(4.6\pm1.5)$ $\times10^{-15}$ and $(0.81\pm0.27)$ $\times10^{-15}$ erg s$^{-1}$ cm$^{-2}$, corresponding to production rate upper limits of $Q$(C$_3$)$\ <(6.9\pm2.3)$ $\times10^{23}$ and $Q$(C$_2$)$\ <(12.2\pm4.1)$ $\times10^{23}$ s$^{-1}$}

\subsection{Brightness profile and Photometry}\label{subsec:profile}

In the previous sections, we showed that the CN emission was activated with 3I/ATLAS inbound at $\sim$3.1 au. Using simultaneous imaging observations with the 1.3\,m telescope, we now examine the evolution of the brightness profile of 3I/ATLAS to explore a possible connection between the onset of gas emission and changes in the comet’s morphology.

\begin{figure}
\plotone{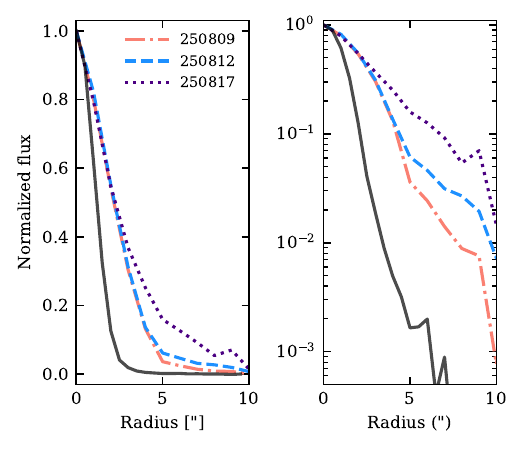}
\caption{Radial profiles of 3I/ATLAS on three different nights. \textbf{Left:} linear scale. \textbf{Right:} logarithmic scale. A common PSF, convolved with the images, is shown for comparison \textcolor{mygreen}{(solid black)}.}
\label{fig:rp}
\end{figure}

Since the seeing conditions varied across nights, we first convolved the comet images to a common PSF. To minimize the effects of elongation along the direction of motion, we used the minimum value of the radial profile. Figure~\ref{fig:rp} shows the radial profiles of 3I/ATLAS on different nights. While the coma brightness increases with time, the difference when the CN emission was only beginning to emerge (August 9 and August 12; see Figure~\ref{fig:CN}) is relatively small. A clearer enhancement of the coma is evident by August 17, coincident with the stronger detection of CN emission.

\begin{figure}
\plotone{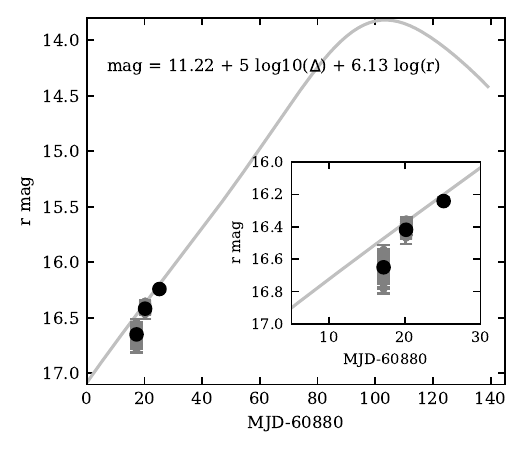}
\caption{Measured brightness and expected brightness evolution of 3I/ATLAS. The expected curve shows the estimated development of the comet’s brightness based on previous observations, fitted to data from the Comet Observations Database (COBS) and the Minor Planet Center (MPC).}
\label{fig:mag}
\end{figure}

On the other hand, the total brightness of the comet follows the expected evolution as a function of its geometry (Figure~\ref{fig:mag}). We did not detect any significant outbursts during our monitoring. As the $U$, $g$, $r$ colors were measured on Aug. 17, we found the comet has $U$-$g$ = $1.7 \pm 0.1$ and $g$-$r$ = $0.62 \pm 0.02$.  The reflectance computed from the 1.3 m photometry is shown in Figure~\ref{fig:reflectance}. It is consistent with the reflectance spectrum obtained by the 2.4 m spectroscope. Again, compared to the spectra obtained in July \citep{Seligman2025, 2025ApJ...990L..27P}, both the photometric and spectroscopic reflectances in August tend to show a deficit on the long wavelength range.

From the photometric data, we calculated the dust production rate parameter $Af\rho$ within a radius of $10^4$ km. We measured $r$-band values of $Af\rho = 260 \pm 20$ cm on August 9, $300 \pm 10$ cm on August 12, and $315 \pm 2$ cm on August 17. The evolution of $Af\rho$ appears correlated with the CN emission, rising steeply during the initial stage and increasing gradually in the subsequent nights. On August 17, we also measured $Af\rho = 272 \pm 2$ cm in the $g$-band and $Af\rho = 180 \pm 20$ cm in the $U$-band. Using the measured color of $g-r = 0.62$, we derived $V-r = 0.195$, yielding a $V$-band value of $Af\rho = 303 \pm 2$ cm on the same night.

\section{Discussion}\label{sec:discussion}

\begin{figure*}[ht]
    \centering
    \includegraphics[width=1\linewidth]{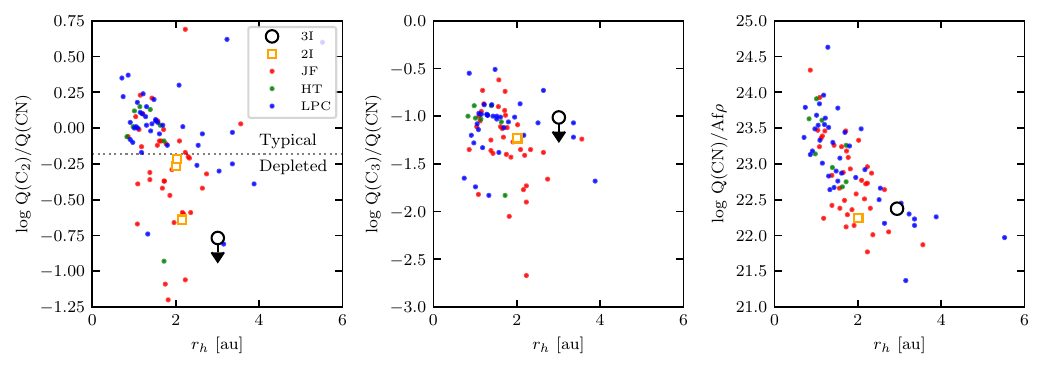}
    \caption{Comparison of carbon-chain to CN ratio and gas-to-dust ratio constraints for interstellar and Solar System comets. \textbf{Left:} Logarithmic C$_2$/CN ratio as a function of heliocentric distance. The \textcolor{mygreen}{white and black circle with a downward-pointing arrow} marks our upper limit for 3I/ATLAS, while the orange squares indicate, in order of decreasing distance, the measurements for 2I/Borisov reported by \citet{Lin2020}, \citet{Bannister2020}, and \citet{Aravind2021}. Points show measurements of Solar System comets from the \citet{2010PDSS.8133E....O} database, with red points corresponding to Jupiter Family Comets (JFC), green points to Halley-type comets (\textcolor{mygreen}{HT}), and blue points to Long-Period Comets (LPC). The horizontal dotted line separates typical from carbon-chain depleted comets. \textbf{Middle:} Logarithmic C$_3$/CN ratio versus heliocentric distance. \textbf{Right:} Logarithmic gas-to-dust ratio versus heliocentric distance. Formatting in the middle and right panels follows the left panel, with only the \citet{Aravind2021} measurement for 2I/Borisov shown. \textcolor{mygreen}{The white and black circle in the right panel represents our measured value for 3I/ATLAS (not an upper limit)}.}    
    \label{fig:production}
\end{figure*}

Both the inferred CN production rates and the heliocentric distance of activation for 3I/ATLAS are relatively typical. For comparison, in the Jupiter-family comet (JFC) 67P/Churyumov–Gerasimenko, an upper limit of $Q$(CN) $\lesssim 4.4\times10^{23}$ s$^{-1}$ was reported between 3.13 and 2.94 au pre-perihelion, while the first confirmed detection occurred much closer to the Sun, at 1.25 au, with $Q$(CN) = $(6.72\pm0.64)\times10^{24}$ s$^{-1}$ \citep{2017MNRAS.469S.222O}. In contrast, comet P/Schwassmann–Wachmann displayed CN emission despite having an almost circular orbit with perihelion at 5.8 au, with a measured $Q$(CN) = $8\times10^{24}$ s$^{-1}$ \citep{1991Icar...90..172C}. While the latter is an extreme case, most reported CN activations in comets occur inside $\sim2.5$ au, and the production rate measured for 3I/ATLAS lies at the lower end of the distribution observed in Solar System comets \citep{1995Icar..118..223A, 2012Icar..218..144C}.

The circumstances surrounding the optical activation of 3I/ATLAS are broadly similar to those of 2I/Borisov, where initially the only detected emission was CN \citep{2019ApJ...885L...9F}. In the case of 2I/Borisov, however, the detection occurred closer to the Sun (2.67 au inbound) and at a similar level, $Q$(CN) $\sim4\times10^{24}$ s$^{-1}$, compared to our measurement for 3I/ATLAS of $\sim6\times10^{24}$ s$^{-1}$ at 3.16 au. Subsequent detections of 2I/Borisov’s CN emission remained relatively stable at $Q$(CN) $\sim2\times10^{24}$ s$^{-1}$ \citep{Opitom2019, Kareta2019, Lin2020, Bannister2020, Aravind2021}.

Although C$_2$ emission was not detected, our measurements already classify 3I/ATLAS as a strongly C$_2$-depleted comet. \citet{1995Icar..118..223A} defined carbon-chain depleted comets as those with a logarithmic C$_2$/CN ratio of $<-0.18$. Using our upper limit for C$_2$ \textcolor{mygreen}{from the coadded spectrum} and the measured CN production rate on August 17, we derive $\log Q(\mathrm{C}_2)/Q(\mathrm{CN}) < \textcolor{mygreen}{(-0.77\pm0.15)}$, which lies at the lower end of values observed in Solar System comets (see left panel of Figure~\ref{fig:production}). For clarity, error bars are omitted from the plot, and note that the heliocentric distances in \citet{2010PDSS.8133E....O} represent mean values. For comparison, 2I/Borisov was also quickly identified as carbon-chain depleted \citep{Opitom2019, Kareta2019}. 

Although the small sample size cautions against broad generalizations, the fact that both known interstellar comets exhibit some degree of carbon-chain depletion may point to a compositional trend among interstellar comets. \textcolor{mygreen}{This trend would resemble that observed in the Solar System, where most carbon-chain depleted comets are found within the Jupiter Family population \citep{1995Icar..118..223A}}. The C$_2$/CN ratio of 2I/Borisov rose rapidly, behavior uncommon among Solar System comets, bringing it closer to the “typical” class. This evolution has been attributed to compositional heterogeneity, with sublimation initially removing a C$_2$-poor surface layer and later exposing material richer in C$_2$ as the comet approached perihelion \citep{Bannister2020, Aravind2021}. Continued monitoring of 3I/ATLAS will be crucial to determine whether C$_2$ emission eventually strengthens. Future observations of interstellar comets will help assess whether carbon-chain depletion is a common property of these visitors, \textcolor{mygreen}{and, if so, whether it originates during their formation or develops later during the journey after ejection from their natal systems}.

In contrast to C$_2$, our current constraints on C$_3$ do not indicate that 3I/ATLAS is particularly depleted in this species (\textcolor{mygreen}{log $Q$(C$_3$)/Q(CN)$\ <(-1.01\pm0.15)$}, see middle panel of Figure~\ref{fig:production}). Our upper limit lies within the range observed for Solar System comets and is larger than the pre-perihelion values measured for 2I/Borisov \citep{Aravind2021}.

Combining the CN production rate with the $V$-band dust production on August 17, we derived a gas-to-dust production ratio. We measure $\log Q(\mathrm{CN})/Af\rho = \textcolor{mygreen}{(22.37\pm0.02)}$, which is comparable to the $22.24\pm0.12$ reported for 2I/Borisov \citep{2019ApJ...885L...9F}, and falls toward the lower end of values in the Lowell Observatory comet database (right panel of Figure~\ref{fig:production}).



\section{Conclusions} \label{sec:conclusion}

In this Letter we have presented the results of a spectroscopic monitoring campaign of 3I/ATLAS with the Hiltner 2.4\,m telescope \textcolor{mygreen}{between August 9-17, during which spectroscopic observations were conducted on five nights. This was} complemented by simultaneous photometric monitoring with the McGraw-Hill 1.3\,m telescope\textcolor{mygreen}{. Both telescopes} operated at the MDM Observatory, during the comet's inbound passage from 3.2 to 2.9\,au. Our main conclusions are:

\begin{itemize}
    \item We detected CN emission at 3870 $\mathrm{\AA}$ in our spectra. It was absent during the first nights of the campaign \textcolor{mygreen}{prior to August 13 UTC,} but became visible \textcolor{mygreen}{thereafter}, with tentative evidence for a steep increase in strength.

    \item The CN molecular production rate \textcolor{mygreen}{on August 16 was measured at $Q$(CN) = $(7.17\pm0.38) \times 10^{24}$ s$^{-1}$}, toward the lower end of values reported for comets.  

    \item Photometric measurements support the spectroscopic detection of gas-driven activity. The brightness profile shows a small but measurable enhancement, consistent with the development of the coma.

    \item The dust production parameter $Af\rho$ was measured as $315 \pm 2$ cm in the $r$-band, $272 \pm 2$ cm in the $g$-band, and $180 \pm 20$ cm in the $U$-band, with values showing a clear increase during the observing period.  

    \item We derived an upper limit of $\log Q(\mathrm{C_2})/Q(\mathrm{CN}) < \textcolor{mygreen}{(-0.77\pm0.15)}$, indicating a strong level of carbon-chain depletion in 3I/ATLAS, according to the compositional classification of \citet{1995Icar..118..223A}. This result may indicate a trend toward carbon-chain depletion in interstellar comets.  

    \item The gas-to-dust production ratio was measured as $\log Q(\mathrm{CN})/Af\rho = \textcolor{mygreen}{(22.37\pm0.02)}$.      
    
\end{itemize}

Taken together, our observations capture the onset of optical gas emission activity in 3I/ATLAS, providing a valuable benchmark in this critical phase and contributing to the broader effort to characterize the object during its Solar System passage. The C$_2$-to-CN production rate constraint falls toward the lower end of values observed in Solar System comets. The emerging similarities with 2I/Borisov and other comets suggest that additional emission features in the optical, such as C$_2$ and C$_3$, may become detectable as 3I/ATLAS continues its approach to the Sun. Continued monitoring from both ground- and space-based facilities will be crucial to further constrain the composition, formation history, and evolutionary pathways of this interstellar visitor \citep{Yaginuma2025,Loeb2025b,Sanchez_2025,Snodgrass2019,Jones2024}.

\begin{acknowledgments}

We gratefully acknowledge the coordination provided by Mario Mateo, Christopher Miller, Jules Halpern, and Eric Galayda, which made the ToO observations reported here possible. L.S.M. and H.W.L. also thank Eric Galayda for his support during the MDM observing run. H.W.L acknowledges Chien-Hsiu Lee for the encouragement and motivation regarding this observation. This work is based on observations obtained at the MDM Observatory, operated by Dartmouth College, Columbia University, Ohio State University, Ohio University, and the University of Michigan. This material is based upon work supported by the National Science Foundation under grant No. AST-2406527. D.Z.S. is supported by an NSF Astronomy and Astrophysics Postdoctoral Fellowship under award AST-2303553. This research award is partially funded by a generous gift of Charles Simonyi to the NSF Division of Astronomical Sciences. The award is made in recognition of significant contributions to Rubin Observatory’s Legacy Survey of Space and Time. \textcolor{mygreen}{We thank the anonymous referee for their comments, which helped improve the quality of this work.}

\end{acknowledgments}





%
\facilities{MDM: Hiltner 2.4 m and McGraw-Hill 1.3 m}

\software{astropy \citep{2013A&A...558A..33A,2018AJ....156..123A,2022ApJ...935..167A}
          }



\bibliography{main}{}

\begin{thebibliography}{}
\expandafter\ifx\csname natexlab\endcsname\relax\def\natexlab#1{#1}\fi
\providecommand{\url}[1]{\href{#1}{#1}}
\providecommand{\dodoi}[1]{doi:~\href{http://doi.org/#1}{\nolinkurl{#1}}}
\providecommand{\doeprint}[1]{\href{http://ascl.net/#1}{\nolinkurl{http://ascl.net/#1}}}
\providecommand{\doarXiv}[1]{\href{https://arxiv.org/abs/#1}{\nolinkurl{https://arxiv.org/abs/#1}}}

\bibitem[{A. {Acker} {et~al.}(1992){Acker}, {Marcout}, {Ochsenbein}, {Stenholm}, {Tylenda}, \& {Schohn}}]{1992secg.book.....A}
{Acker}, A., {Marcout}, J., {Ochsenbein}, F., {et~al.} 1992, {The Strasbourg-ESO Catalogue of Galactic Planetary Nebulae. Parts I, II.}

\bibitem[{M.~F. {A'Hearn} {et~al.}(1995){A'Hearn}, {Millis}, {Schleicher}, {Osip}, \& {Birch}}]{1995Icar..118..223A}
{A'Hearn}, M.~F., {Millis}, R.~C., {Schleicher}, D.~O., {Osip}, D.~J., \& {Birch}, P.~V. 1995, \bibinfo{title}{{The ensemble properties of comets: Results from narrowband photometry of 85 comets, 1976-1992.},} \icarus, 118, 223, \dodoi{10.1006/icar.1995.1190}

\bibitem[{M.~R. {Alarcon} {et~al.}(2025){Alarcon}, {Serra-Ricart}, {Licandro}, {Arencibia}, {Ruiz Cejudo}, \& {Trujillo}}]{2025ATel17264....1A}
{Alarcon}, M.~R., {Serra-Ricart}, M., {Licandro}, J., {et~al.} 2025, \bibinfo{title}{{Deep g'-band Imaging of Interstellar Comet 3I/ATLAS from the Two-meter Twin Telescope (TTT)},} The Astronomer's Telegram, 17264, 1

\bibitem[{A. {Alvarez-Candal} {et~al.}(2025){Alvarez-Candal}, {Rizos}, {Lara}, {Santos-Sanz}, {Gutierrez}, {Ortiz}, \& {Morales}}]{2025A&A...700L..10A}
{Alvarez-Candal}, A., {Rizos}, J.~L., {Lara}, L.~M., {et~al.} 2025, \bibinfo{title}{{X-SHOOTER spectrum of comet 3I/ATLAS: Insights into a distant interstellar visitor},} \aap, 700, L10, \dodoi{10.1051/0004-6361/202556338}

\bibitem[{K. {Aravind} {et~al.}(2021){Aravind}, {Ganesh}, {Venkataramani}, {Sahu}, {Angchuk}, {Sivarani}, \& {Unni}}]{Aravind2021}
{Aravind}, K., {Ganesh}, S., {Venkataramani}, K., {et~al.} 2021, \bibinfo{title}{{Activity of the first interstellar comet 2I/Borisov around perihelion: results from Indian observatories},} \mnras, 502, 3491, \dodoi{10.1093/mnras/stab084}

\bibitem[{ {Astropy Collaboration} {et~al.}(2013){Astropy Collaboration}, {Robitaille}, {Tollerud}, {Greenfield}, {Droettboom}, {Bray}, {Aldcroft}, {Davis}, {Ginsburg}, {Price-Whelan}, {Kerzendorf}, {Conley}, {Crighton}, {Barbary}, {Muna}, {Ferguson}, {Grollier}, {Parikh}, {Nair}, {Unther}, {Deil}, {Woillez}, {Conseil}, {Kramer}, {Turner}, {Singer}, {Fox}, {Weaver}, {Zabalza}, {Edwards}, {Azalee Bostroem}, {Burke}, {Casey}, {Crawford}, {Dencheva}, {Ely}, {Jenness}, {Labrie}, {Lim}, {Pierfederici}, {Pontzen}, {Ptak}, {Refsdal}, {Servillat}, \& {Streicher}}]{2013A&A...558A..33A}
{Astropy Collaboration}, {Robitaille}, T.~P., {Tollerud}, E.~J., {et~al.} 2013, \bibinfo{title}{{Astropy: A community Python package for astronomy},} \aap, 558, A33, \dodoi{10.1051/0004-6361/201322068}

\bibitem[{ {Astropy Collaboration} {et~al.}(2018){Astropy Collaboration}, {Price-Whelan}, {Sip{\H{o}}cz}, {G{\"u}nther}, {Lim}, {Crawford}, {Conseil}, {Shupe}, {Craig}, {Dencheva}, {Ginsburg}, {VanderPlas}, {Bradley}, {P{\'e}rez-Su{\'a}rez}, {de Val-Borro}, {Aldcroft}, {Cruz}, {Robitaille}, {Tollerud}, {Ardelean}, {Babej}, {Bach}, {Bachetti}, {Bakanov}, {Bamford}, {Barentsen}, {Barmby}, {Baumbach}, {Berry}, {Biscani}, {Boquien}, {Bostroem}, {Bouma}, {Brammer}, {Bray}, {Breytenbach}, {Buddelmeijer}, {Burke}, {Calderone}, {Cano Rodr{\'\i}guez}, {Cara}, {Cardoso}, {Cheedella}, {Copin}, {Corrales}, {Crichton}, {D'Avella}, {Deil}, {Depagne}, {Dietrich}, {Donath}, {Droettboom}, {Earl}, {Erben}, {Fabbro}, {Ferreira}, {Finethy}, {Fox}, {Garrison}, {Gibbons}, {Goldstein}, {Gommers}, {Greco}, {Greenfield}, {Groener}, {Grollier}, {Hagen}, {Hirst}, {Homeier}, {Horton}, {Hosseinzadeh}, {Hu}, {Hunkeler}, {Ivezi{\'c}}, {Jain}, {Jenness}, {Kanarek}, {Kendrew}, {Kern}, {Kerzendorf}, {Khvalko}, {King}, {Kirkby}, {Kulkarni},
  {Kumar}, {Lee}, {Lenz}, {Littlefair}, {Ma}, {Macleod}, {Mastropietro}, {McCully}, {Montagnac}, {Morris}, {Mueller}, {Mumford}, {Muna}, {Murphy}, {Nelson}, {Nguyen}, {Ninan}, {N{\"o}the}, {Ogaz}, {Oh}, {Parejko}, {Parley}, {Pascual}, {Patil}, {Patil}, {Plunkett}, {Prochaska}, {Rastogi}, {Reddy Janga}, {Sabater}, {Sakurikar}, {Seifert}, {Sherbert}, {Sherwood-Taylor}, {Shih}, {Sick}, {Silbiger}, {Singanamalla}, {Singer}, {Sladen}, {Sooley}, {Sornarajah}, {Streicher}, {Teuben}, {Thomas}, {Tremblay}, {Turner}, {Terr{\'o}n}, {van Kerkwijk}, {de la Vega}, {Watkins}, {Weaver}, {Whitmore}, {Woillez}, {Zabalza}, \& {Astropy Contributors}}]{2018AJ....156..123A}
{Astropy Collaboration}, {Price-Whelan}, A.~M., {Sip{\H{o}}cz}, B.~M., {et~al.} 2018, \bibinfo{title}{{The Astropy Project: Building an Open-science Project and Status of the v2.0 Core Package},} \aj, 156, 123, \dodoi{10.3847/1538-3881/aabc4f}

\bibitem[{ {Astropy Collaboration} {et~al.}(2022){Astropy Collaboration}, {Price-Whelan}, {Lim}, {Earl}, {Starkman}, {Bradley}, {Shupe}, {Patil}, {Corrales}, {Brasseur}, {N{\"o}the}, {Donath}, {Tollerud}, {Morris}, {Ginsburg}, {Vaher}, {Weaver}, {Tocknell}, {Jamieson}, {van Kerkwijk}, {Robitaille}, {Merry}, {Bachetti}, {G{\"u}nther}, {Aldcroft}, {Alvarado-Montes}, {Archibald}, {B{\'o}di}, {Bapat}, {Barentsen}, {Baz{\'a}n}, {Biswas}, {Boquien}, {Burke}, {Cara}, {Cara}, {Conroy}, {Conseil}, {Craig}, {Cross}, {Cruz}, {D'Eugenio}, {Dencheva}, {Devillepoix}, {Dietrich}, {Eigenbrot}, {Erben}, {Ferreira}, {Foreman-Mackey}, {Fox}, {Freij}, {Garg}, {Geda}, {Glattly}, {Gondhalekar}, {Gordon}, {Grant}, {Greenfield}, {Groener}, {Guest}, {Gurovich}, {Handberg}, {Hart}, {Hatfield-Dodds}, {Homeier}, {Hosseinzadeh}, {Jenness}, {Jones}, {Joseph}, {Kalmbach}, {Karamehmetoglu}, {Ka{\l}uszy{\'n}ski}, {Kelley}, {Kern}, {Kerzendorf}, {Koch}, {Kulumani}, {Lee}, {Ly}, {Ma}, {MacBride}, {Maljaars}, {Muna}, {Murphy}, {Norman},
  {O'Steen}, {Oman}, {Pacifici}, {Pascual}, {Pascual-Granado}, {Patil}, {Perren}, {Pickering}, {Rastogi}, {Roulston}, {Ryan}, {Rykoff}, {Sabater}, {Sakurikar}, {Salgado}, {Sanghi}, {Saunders}, {Savchenko}, {Schwardt}, {Seifert-Eckert}, {Shih}, {Jain}, {Shukla}, {Sick}, {Simpson}, {Singanamalla}, {Singer}, {Singhal}, {Sinha}, {Sip{\H{o}}cz}, {Spitler}, {Stansby}, {Streicher}, {{\v{S}}umak}, {Swinbank}, {Taranu}, {Tewary}, {Tremblay}, {de Val-Borro}, {Van Kooten}, {Vasovi{\'c}}, {Verma}, {de Miranda Cardoso}, {Williams}, {Wilson}, {Winkel}, {Wood-Vasey}, {Xue}, {Yoachim}, {Zhang}, {Zonca}, \& {Astropy Project Contributors}}]{2022ApJ...935..167A}
{Astropy Collaboration}, {Price-Whelan}, A.~M., {Lim}, P.~L., {et~al.} 2022, \bibinfo{title}{{The Astropy Project: Sustaining and Growing a Community-oriented Open-source Project and the Latest Major Release (v5.0) of the Core Package},} \apj, 935, 167, \dodoi{10.3847/1538-4357/ac7c74}

\bibitem[{M.~T. {Bannister} {et~al.}(2020){Bannister}, {Opitom}, {Fitzsimmons}, {Moulane}, {Jehin}, {Seligman}, {Rousselot}, {Knight}, {Marsset}, {Schwamb}, {Guilbert-Lepoutre}, {Jorda}, {Vernazza}, \& {Benkhaldoun}}]{Bannister2020}
{Bannister}, M.~T., {Opitom}, C., {Fitzsimmons}, A., {et~al.} 2020, \bibinfo{title}{{Interstellar comet 2I/Borisov as seen by MUSE: C$_2$, NH$_2$ and red CN detections},} arXiv e-prints, arXiv:2001.11605.
\newblock \doarXiv{2001.11605}

\bibitem[{M. {Belyakov} {et~al.}(2025){Belyakov}, {Fremling}, {Graham}, {Bolin}, {Kilic}, {Jewett}, {Lisse}, {Ingebretsen}, {Davis}, \& {Wong}}]{2025RNAAS...9..194B}
{Belyakov}, M., {Fremling}, C., {Graham}, M.~J., {et~al.} 2025, \bibinfo{title}{{Palomar and Apache Point Spectrophotometry of Interstellar Comet 3I/ATLAS},} Research Notes of the American Astronomical Society, 9, 194, \dodoi{10.3847/2515-5172/adf059}

\bibitem[{J. {Beniyama}(2025){Beniyama}}]{Beniyama2025}
{Beniyama}, J. 2025, \bibinfo{title}{{Simultaneous visible spectrophotometry of interstellar object 3I/ATLAS with Seimei/TriCCS},} \pasj, \dodoi{10.1093/pasj/psaf097}

\bibitem[{J.~B. {Bergner} \& D.~Z. {Seligman}(2023){Bergner} \& {Seligman}}]{Bergner2023}
{Bergner}, J.~B., \& {Seligman}, D.~Z. 2023, \bibinfo{title}{{Acceleration of 1I/`Oumuamua from radiolytically produced H$_{2}$ in H$_{2}$O ice},} \nat, 615, 610, \dodoi{10.1038/s41586-022-05687-w}

\bibitem[{M. {Bessell} {et~al.}(2011){Bessell}, {Bloxham}, {Schmidt}, {Keller}, {Tisserand}, \& {Francis}}]{skymapper_filters}
{Bessell}, M., {Bloxham}, G., {Schmidt}, B., {et~al.} 2011, \bibinfo{title}{{SkyMapper Filter Set: Design and Fabrication of Large-Scale Optical Filters},} \pasp, 123, 789, \dodoi{10.1086/660849}

\bibitem[{D. {Bockel{\'e}e-Morvan} \& N. {Biver}(2017){Bockel{\'e}e-Morvan} \& {Biver}}]{2017RSPTA.37560252B}
{Bockel{\'e}e-Morvan}, D., \& {Biver}, N. 2017, \bibinfo{title}{{The composition of cometary ices},} Philosophical Transactions of the Royal Society of London Series A, 375, 20160252, \dodoi{10.1098/rsta.2016.0252}

\bibitem[{D. {Bodewits} {et~al.}(2020){Bodewits}, {Noonan}, {Feldman}, {Bannister}, {Farnocchia}, {Harris}, {Li}, {Mandt}, {Parker}, \& {Xing}}]{Bodewits2020}
{Bodewits}, D., {Noonan}, J.~W., {Feldman}, P.~D., {et~al.} 2020, \bibinfo{title}{{The carbon monoxide-rich interstellar comet 2I/Borisov},} Nature Astronomy, 4, 867, \dodoi{10.1038/s41550-020-1095-2}

\bibitem[{G. {Borisov} {et~al.}(2019){Borisov}, {Durig}, {Sato}, {Birtwhistle}, {Chen}, {Green}, {Bacci}, {Maestripieri}, \& {Nakano}}]{borisov_2I_cbet}
{Borisov}, G., {Durig}, D.~T., {Sato}, H., {et~al.} 2019, \bibinfo{title}{{Comet C/2019 Q4 (Borisov)},} Central Bureau Electronic Telegrams, 4666, 1

\bibitem[{C.~O. {Chandler} {et~al.}(2025){Chandler}, {Bernardinelli}, {Juri{\'c}}, {Singh}, {Hsieh}, {Sullivan}, {Jones}, {Kurlander}, {Vavilov}, {Eggl}, {Holman}, {Spoto}, {Schwamb}, {Christensen}, {Beebe}, {Roodman}, {Lim}, {Jenness}, {Bosch}, {Smart}, {Bellm}, {MacBride}, {Rawls}, {Greenstreet}, {Slater}, {Heinze}, {Ivezi{\'c}}, {Blum}, {Connolly}, {Daues}, {Makadia}, {Gower}, {Bryce Kalmbach}, {Monet}, {Bannister}, {Dones}, {Dorsey}, {Fraser}, {Forbes}, {Fuentes}, {Holt}, {Inno}, {Jones}, {Knight}, {Lintott}, {Lister}, {Lupton}, {Mendoza Magbanua}, {Malhotra}, {Mueller}, {Murtagh}, {Pandey}, {Reach}, {Samarasinha}, {Seligman}, {Snodgrass}, {Solontoi}, {Szab{\'o}}, {White}, {Womack}, {Young}, {Allbery}, {Armellin}, {Aubourg}, {Avdellidou}, {Azfar}, {Bauer}, {Bechtol}, {Belyakov}, {Benecchi}, {Bertini}, {Bolin}, {Bose}, {Buchanan}, {Boucaud}, {Boufleur}, {Boutigny}, {Braga-Ribas}, {Calabrese}, {Camargo}, {Caplar}, {Carry}, {Carvajal}, {Choi}, {Cowan}, {Croft}, {{\'C}uk}, {Daruich}, {Daubard}, {Davenport},
  {Daylan}, {Delgado}, {Devillepoix}, {Doherty}, {Donaldson}, {Drass}, {Deppe}, {Dubois-Felsmann}, {Economou}, {Eduardo}, {Farnocchia}, {Frissell}, {Fedorets}, {Fernandes}, {Fulle}, {Gerdes}, {Gibbs}, {Gillan}, {Guy}, {Hammergren}, {Hanushevsky}, {Hernandez}, {Hestroffer}, {Hopkins}, {Granvik}, {Ieva}, {Irving}, {Jannuzi}, {Jimenez}, {Ramos Gomes-J{\'u}nior}, {Juramy}, {Kahn}, {Kannawadi}, {Kang}, {Kryszczy{\'n}ska}, {Kotov}, {Koumjian}, {Krughoff}, {Lage}, {Lange}, {Levine}, {Li}, {Licandro}, {Lin}, {Lust}, {Lyttle}, {Mahabal}, {Mahlke}, {Plazas Malag{\'o}n}, {Salazar Manzano}, {Marc}, {Margoti}, {Mar{\v{c}}eta}, {Menanteau}, {Meyers}, {Mills}, {Morato}, {More}, {Morrison}, {Moulane}, {Mu{\~n}oz-Guti{\'e}rrez}, {M.}, {O'Connor}, {Oldag}, {Oldroyd}, {O'Mullane}, {Opitom}, {Oszkiewicz}, {Page}, {Patterson}, {Payne}, {Peloton}, {Pereira}, {Peterson}, {Polin}, {Pollek}, {Polen}, {Qiu}, {Ragozzine}, {Rajagopal}, {van Reeven}, {Rice}, {Ridgway}, {Rivkin}, {Robinson}, {Ro{\.z}ek}, {Salnikov}, {S{\'a}nchez},
  {Sarid}, {Schambeau}, {Scolnic}, {Schindler}, {Seaman}, {Jacques}, {Shaw}, {Shugart}, {Sick}, {Siraj}, {Sitarz}, {Sobhani}, {Soldahl}, {Stalder}, {Stetzler}, {Swinbank}, {Szigeti}, {Tauraso}, {Thornton}, {Tonietti}, {Trilling}, \& {Trujillo}}]{Chandler2025}
{Chandler}, C.~O., {Bernardinelli}, P.~H., {Juri{\'c}}, M., {et~al.} 2025, \bibinfo{title}{{NSF-DOE Vera C. Rubin Observatory Observations of Interstellar Comet 3I/ATLAS (C/2025 N1)},} arXiv e-prints, arXiv:2507.13409.
\newblock \doarXiv{2507.13409}

\bibitem[{A.~L. {Cochran} {et~al.}(2012){Cochran}, {Barker}, \& {Gray}}]{2012Icar..218..144C}
{Cochran}, A.~L., {Barker}, E.~S., \& {Gray}, C.~L. 2012, \bibinfo{title}{{Thirty years of cometary spectroscopy from McDonald Observatory},} \icarus, 218, 144, \dodoi{10.1016/j.icarus.2011.12.010}

\bibitem[{A.~L. {Cochran} \& W.~D. {Cochran}(1991){Cochran} \& {Cochran}}]{1991Icar...90..172C}
{Cochran}, A.~L., \& {Cochran}, W.~D. 1991, \bibinfo{title}{{The first detection of CN and the distribution of CO $^{+}$ gas in the coma of Comet P/Schwassmann-Wachman 1},} \icarus, 90, 172, \dodoi{10.1016/0019-1035(91)90077-7}

\bibitem[{A.~L. {Cochran} {et~al.}(2015){Cochran}, {Levasseur-Regourd}, {Cordiner}, {Hadamcik}, {Lasue}, {Gicquel}, {Schleicher}, {Charnley}, {Mumma}, {Paganini}, {Bockel{\'e}e-Morvan}, {Biver}, \& {Kuan}}]{2015SSRv..197....9C}
{Cochran}, A.~L., {Levasseur-Regourd}, A.-C., {Cordiner}, M., {et~al.} 2015, \bibinfo{title}{{The Composition of Comets},} \ssr, 197, 9, \dodoi{10.1007/s11214-015-0183-6}

\bibitem[{M.~R. {Combi} {et~al.}(2004){Combi}, {Harris}, \& {Smyth}}]{2004come.book..523C}
{Combi}, M.~R., {Harris}, W.~M., \& {Smyth}, W.~H. 2004, in Comets II, ed. M.~C. {Festou}, H.~U. {Keller}, \& H.~A. {Weaver}, 523

\bibitem[{M.~A. {Cordiner} {et~al.}(2020){Cordiner}, {Milam}, {Biver}, {Bockel{\'e}e-Morvan}, {Roth}, {Bergin}, {Jehin}, {Remijan}, {Charnley}, {Mumma}, {Boissier}, {Crovisier}, {Paganini}, {Kuan}, \& {Lis}}]{Cordiner2020}
{Cordiner}, M.~A., {Milam}, S.~N., {Biver}, N., {et~al.} 2020, \bibinfo{title}{{Unusually high CO abundance of the first active interstellar comet},} Nature Astronomy, 4, 861, \dodoi{10.1038/s41550-020-1087-2}

\bibitem[{M.~A. {Cordiner} {et~al.}(2025){Cordiner}, {Roth}, {Kelley}, {Bodewits}, {Charnley}, {Drozdovskaya}, {Farnocchia}, {Micheli}, {Milam}, {Opitom}, {Schwamb}, \& {Thomas}}]{Cordiner2025}
{Cordiner}, M.~A., {Roth}, N.~X., {Kelley}, M. S.~P., {et~al.} 2025, \bibinfo{title}{{JWST detection of a carbon dioxide dominated gas coma surrounding interstellar object 3I/ATLAS},} arXiv e-prints, arXiv:2508.18209, \dodoi{10.48550/arXiv.2508.18209}

\bibitem[{R. {de la Fuente Marcos} {et~al.}(2025){de la Fuente Marcos}, {Alarcon}, {Licandro}, {Serra-Ricart}, {de Le{\'o}n}, {de la Fuente Marcos}, {Lombardi}, {Tejero}, {Cabrera-Lavers}, {Guerra Arencibia}, \& {Ruiz Cejudo}}]{2025A&A...700L...9D}
{de la Fuente Marcos}, R., {Alarcon}, M.~R., {Licandro}, J., {et~al.} 2025, \bibinfo{title}{{Assessing interstellar comet 3I/ATLAS with the 10.4 m Gran Telescopio Canarias and the Two-meter Twin Telescope},} \aap, 700, L9, \dodoi{10.1051/0004-6361/202556439}

\bibitem[{F.~E. {DeMeo} {et~al.}(2009){DeMeo}, {Binzel}, {Slivan}, \& {Bus}}]{2009Icar..202..160D}
{DeMeo}, F.~E., {Binzel}, R.~P., {Slivan}, S.~M., \& {Bus}, S.~J. 2009, \bibinfo{title}{{An extension of the Bus asteroid taxonomy into the near-infrared},} \icarus, 202, 160, \dodoi{10.1016/j.icarus.2009.02.005}

\bibitem[{L. {Denneau} {et~al.}(2025){Denneau}, {Siverd}, {Tonry}, {Weiland}, {Erasmus}, {Fitzsimmons}, \& {Robinson}}]{Denneau2025}
{Denneau}, L., {Siverd}, R., {Tonry}, J., {et~al.} 2025, \bibinfo{title}{{3I/ATLAS = C/2025 N1 (ATLAS)},} MPEC

\bibitem[{S.~J. Desch \& A.~P. Jackson(2021)Desch \& Jackson}]{desch20211i}
Desch, S.~J., \& Jackson, A.~P. 2021, \bibinfo{title}{1I/‘Oumuamua as an N2 ice fragment of an exo-pluto surface II: Generation of N2 ice fragments and the origin of ‘Oumuamua,} Journal of Geophysical Research: Planets, e2020JE006807

\bibitem[{S.~J. {Desch} \& A.~P. {Jackson}(2022){Desch} \& {Jackson}}]{Desch2022}
{Desch}, S.~J., \& {Jackson}, A.~P. 2022, \bibinfo{title}{{Some Pertinent Issues for Interstellar Panspermia Raised after the Discovery of 1I/`Oumuamua},} Astrobiology, 22, 1400, \dodoi{10.1089/ast.2021.0199}

\bibitem[{L. {Dones} {et~al.}(2015){Dones}, {Brasser}, {Kaib}, \& {Rickman}}]{2015SSRv..197..191D}
{Dones}, L., {Brasser}, R., {Kaib}, N., \& {Rickman}, H. 2015, \bibinfo{title}{{Origin and Evolution of the Cometary Reservoirs},} \ssr, 197, 191, \dodoi{10.1007/s11214-015-0223-2}

\bibitem[{T.~L. {Farnham} {et~al.}(2000){Farnham}, {Schleicher}, \& {A'Hearn}}]{2000Icar..147..180F}
{Farnham}, T.~L., {Schleicher}, D.~G., \& {A'Hearn}, M.~F. 2000, \bibinfo{title}{{The HB Narrowband Comet Filters: Standard Stars and Calibrations},} \icarus, 147, 180, \dodoi{10.1006/icar.2000.6420}

\bibitem[{D. {Farnocchia} {et~al.}(2023){Farnocchia}, {Seligman}, {Granvik}, {Hainaut}, {Meech}, {Micheli}, {Weryk}, {Chesley}, {Christensen}, {Koschny}, {Kleyna}, {Lazzaro}, {Mommert}, \& {Wainscoat}}]{Farnocchia2023}
{Farnocchia}, D., {Seligman}, D.~Z., {Granvik}, M., {et~al.} 2023, \bibinfo{title}{{(523599) 2003 RM: The Asteroid that Wanted to be a Comet},} \psj, 4, 29, \dodoi{10.3847/PSJ/acb25b}

\bibitem[{A.~D. {Feinstein} {et~al.}(2025){Feinstein}, {Noonan}, \& {Seligman}}]{Feinstein2025}
{Feinstein}, A.~D., {Noonan}, J.~W., \& {Seligman}, D.~Z. 2025, \bibinfo{title}{{Precovery Observations of 3I/ATLAS from TESS Suggest Possible Distant Activity},} \apjl, 991, L2, \dodoi{10.3847/2041-8213/adfd4d}

\bibitem[{P.~D. {Feldman} {et~al.}(2004){Feldman}, {Cochran}, \& {Combi}}]{2004come.book..425F}
{Feldman}, P.~D., {Cochran}, A.~L., \& {Combi}, M.~R. 2004, in Comets II, ed. M.~C. {Festou}, H.~U. {Keller}, \& H.~A. {Weaver}, 425

\bibitem[{A. {Fitzsimmons} {et~al.}(2024){Fitzsimmons}, {Meech}, {Matr{\`a}}, \& {Pfalzner}}]{Fitzsimmons2024}
{Fitzsimmons}, A., {Meech}, K., {Matr{\`a}}, L., \& {Pfalzner}, S. 2024, in Comets III, ed. K.~J. {Meech}, M.~R. {Combi}, D.~{Bockel{\'e}e-Morvan}, S.~N. {Raymodn}, \& M.~E. {Zolensky}, 731--766

\bibitem[{A. {Fitzsimmons} {et~al.}(2019){Fitzsimmons}, {Hainaut}, {Meech}, {Jehin}, {Moulane}, {Opitom}, {Yang}, {Keane}, {Kleyna}, {Micheli}, \& {Snodgrass}}]{2019ApJ...885L...9F}
{Fitzsimmons}, A., {Hainaut}, O., {Meech}, K.~J., {et~al.} 2019, \bibinfo{title}{{Detection of CN Gas in Interstellar Object 2I/Borisov},} \apjl, 885, L9, \dodoi{10.3847/2041-8213/ab49fc}

\bibitem[{E.~G. {Flekk{\o}y} {et~al.}(2019){Flekk{\o}y}, {Luu}, \& {Toussaint}}]{Flekkoy19}
{Flekk{\o}y}, E.~G., {Luu}, J., \& {Toussaint}, R. 2019, \bibinfo{title}{{The Interstellar Object {\textquoteright}Oumuamua as a Fractal Dust Aggregate},} \apjl, 885, L41, \dodoi{10.3847/2041-8213/ab4f78}

\bibitem[{N. {Fray} {et~al.}(2005){Fray}, {B{\'e}nilan}, {Cottin}, {Gazeau}, \& {Crovisier}}]{2005P&SS...53.1243F}
{Fray}, N., {B{\'e}nilan}, Y., {Cottin}, H., {Gazeau}, M.~C., \& {Crovisier}, J. 2005, \bibinfo{title}{{The origin of the CN radical in comets: A review from observations and models},} \planss, 53, 1243, \dodoi{10.1016/j.pss.2005.06.005}

\bibitem[{A. {Guilbert-Lepoutre} {et~al.}(2015){Guilbert-Lepoutre}, {Besse}, {Mousis}, {Ali-Dib}, {H{\"o}fner}, {Koschny}, \& {Hager}}]{2015SSRv..197..271G}
{Guilbert-Lepoutre}, A., {Besse}, S., {Mousis}, O., {et~al.} 2015, \bibinfo{title}{{On the Evolution of Comets},} \ssr, 197, 271, \dodoi{10.1007/s11214-015-0148-9}

\bibitem[{L. {Haser}(1957){Haser}}]{1957BSRSL..43..740H}
{Haser}, L. 1957, \bibinfo{title}{{Distribution d'intensit{\'e} dans la t{\^e}te d'une com{\`e}te},} Bulletin de la Societe Royale des Sciences de Liege, 43, 740

\bibitem[{L. {Haser} {et~al.}(2020){Haser}, {Oset}, \& {Bodewits}}]{2020PSJ.....1...83H}
{Haser}, L., {Oset}, S., \& {Bodewits}, D. 2020, \bibinfo{title}{{Intensity Distribution in the Heads of Comets},} \psj, 1, 83, \dodoi{10.3847/PSJ/abc17b}

\bibitem[{J. {Helbert} {et~al.}(2005){Helbert}, {Rauer}, {Boice}, \& {Huebner}}]{2005A&A...442.1107H}
{Helbert}, J., {Rauer}, H., {Boice}, D.~C., \& {Huebner}, W.~F. 2005, \bibinfo{title}{{The chemistry of C$_{2}$ and C$_{3}$ in the coma of Comet C/1995 O1 (Hale-Bopp) at heliocentric distances r$_{h}$ {\ensuremath{\geq}} 2.9 AU},} \aap, 442, 1107, \dodoi{10.1051/0004-6361:20041571}

\bibitem[{M.~J. {Hopkins} {et~al.}(2025){Hopkins}, {Dorsey}, {Forbes}, {Bannister}, {Lintott}, \& {Leicester}}]{2025arXiv250705318H}
{Hopkins}, M.~J., {Dorsey}, R.~C., {Forbes}, J.~C., {et~al.} 2025, \bibinfo{title}{{From a Different Star: 3I/ATLAS in the context of the {\={O}}tautahi-Oxford interstellar object population model},} arXiv e-prints, arXiv:2507.05318, \dodoi{10.48550/arXiv.2507.05318}

\bibitem[{C.~A. {Ihalawela} {et~al.}(2011){Ihalawela}, {Pierce}, {Dorman}, \& {Cochran}}]{2011ApJ...741...89I}
{Ihalawela}, C.~A., {Pierce}, D.~M., {Dorman}, G.~R., \& {Cochran}, A.~L. 2011, \bibinfo{title}{{The Spatial Distribution of OH and CN Radicals in the Coma of Comet Encke},} \apj, 741, 89, \dodoi{10.1088/0004-637X/741/2/89}

\bibitem[{A.~P. Jackson \& S.~J. Desch(2021)Jackson \& Desch}]{jackson20211i}
Jackson, A.~P., \& Desch, S.~J. 2021, \bibinfo{title}{1I/‘Oumuamua as an N2 ice fragment of an exo-Pluto surface: I. Size and Compositional Constraints,} Journal of Geophysical Research: Planets, e2020JE006706

\bibitem[{D. {Jewitt} {et~al.}(2025){Jewitt}, {Hui}, {Mutchler}, {Kim}, \& {Agarwal}}]{2025arXiv250802934J}
{Jewitt}, D., {Hui}, M.-T., {Mutchler}, M., {Kim}, Y., \& {Agarwal}, J. 2025, \bibinfo{title}{{Hubble Space Telescope Observations of the Interstellar Interloper 3I/ATLAS},} arXiv e-prints, arXiv:2508.02934, \dodoi{10.48550/arXiv.2508.02934}

\bibitem[{D. {Jewitt} \& J. {Luu}(2025){Jewitt} \& {Luu}}]{2025ATel17263....1J}
{Jewitt}, D., \& {Luu}, J. 2025, \bibinfo{title}{{Interstellar Interloper C/2025 N1 is Active},} The Astronomer's Telegram, 17263, 1

\bibitem[{D. {Jewitt} {et~al.}(2017){Jewitt}, {Luu}, {Rajagopal}, {Kotulla}, {Ridgway}, {Liu}, \& {Augusteijn}}]{Jewitt2017}
{Jewitt}, D., {Luu}, J., {Rajagopal}, J., {et~al.} 2017, \bibinfo{title}{{Interstellar Interloper 1I/2017 U1: Observations from the NOT and WIYN Telescopes},} Astrophysical Journal Letters, 850, L36, \dodoi{10.3847/2041-8213/aa9b2f}

\bibitem[{D. {Jewitt} \& D.~Z. {Seligman}(2023){Jewitt} \& {Seligman}}]{Jewitt2023ARAA}
{Jewitt}, D., \& {Seligman}, D.~Z. 2023, \bibinfo{title}{{The Interstellar Interlopers},} \araa, 61, 197, \dodoi{10.1146/annurev-astro-071221-054221}

\bibitem[{G.~H. Jones {et~al.}(2024)Jones, Snodgrass, Tubiana, K\"{u}ppers, Kawakita, Lara, Agarwal, André, Attree, Auster, Bagnulo, Bannister, Beth, Bowles, Coates, Colangeli, Corral van Damme, Da~Deppo, De~Keyser, Della~Corte, Edberg, El-Maarry, Faggi, Fulle, Funase, Galand, Goetz, Groussin, Guilbert-Lepoutre, Henri, Kasahara, Kereszturi, Kidger, Knight, Kokotanekova, Kolmasova, Kossacki, K\"{u}hrt, Kwon, La~Forgia, Levasseur-Regourd, Lippi, Longobardo, Marschall, Morawski, Muñoz, N\"{a}sil\"{a}, Nilsson, Opitom, Pajusalu, Pommerol, Prech, Rando, Ratti, Rothkaehl, Rotundi, Rubin, Sakatani, Sánchez, Simon~Wedlund, Stankov, Thomas, Toth, Villanueva, Vincent, Volwerk, Wurz, Wielders, Yoshioka, Aleksiejuk, Alvarez, Amoros, Aslam, Atamaniuk, Baran, Barciński, Beck, Behnke, Berglund, Bertini, Bieda, Binczyk, Busch, Cacovean, Capria, Carr, Castro~Marín, Ceriotti, Chioetto, Chuchra-Konrad, Cocola, Colin, Crews, Cripps, Cupido, Dassatti, Davidsson, De~Roche, Deca, Del~Togno, Dhooghe, Donaldson~Hanna,
  Eriksson, Fedorov, Fernández-Valenzuela, Ferretti, Floriot, Frassetto, Fredriksson, Garnier, Gaweł, Génot, Gerber, Glassmeier, Granvik, Grison, Gunell, Hachemi, Hagen, Hajra, Harada, Hasiba, Haslebacher, Herranz De La Revilla, Hestroffer, Hewagama, Holt, Hviid, Iakubivskyi, Inno, Irwin, Ivanovski, Jansky, Jernej, Jeszenszky, Jimenéz, Jorda, Kama, Kameda, Kelley, Klepacki, Kohout, Kojima, Kowalski, Kuwabara, Ladno, Laky, Lammer, Lan, Lavraud, Lazzarin, Le~Duff, Lee, Lesniak, Lewis, Lin, Lister, Lowry, Magnes, Markkanen, Martinez~Navajas, Martins, Matsuoka, Matyjasiak, Mazelle, Mazzotta Epifani, Meier, Michaelis, Micheli, Migliorini, Millet, Moreno, Mottola, Moutounaick, Muinonen, M\"{u}ller, Murakami, Murata, Myszka, Nakajima, Nemeth, Nikolajev, Nordera, Ohlsson, Olesk, Ottacher, Ozaki, Oziol, Patel, Savio~Paul, Penttil\"{a}, Pernechele, Peterson, Petraglio, Piccirillo, Plaschke, Polak, Postberg, Proosa, Protopapa, Puccio, Ranvier, Raymond, Richter, Rieder, Rigamonti, Ruiz~Rodriguez, Santolik,
  Sasaki, Schr\"{o}dter, Shirley, Slavinskis, Sodor, Soucek, Stephenson, St\"{o}ckli, Szewczyk, Troznai, Uhlir, Usami, Valavanoglou, Vaverka, Wang, Wang, Wattieaux, Wieser, Wolf, Yano, Yoshikawa, Zakharov, Zawistowski, Zuppella, Rinaldi, \& Ji}]{Jones2024}
Jones, G.~H., Snodgrass, C., Tubiana, C., {et~al.} 2024, \bibinfo{title}{The Comet Interceptor Mission,} Space Science Reviews, 220, \dodoi{10.1007/s11214-023-01035-0}

\bibitem[{T. {Kareta} {et~al.}(2020){Kareta}, {Andrews}, {Noonan}, {Harris}, {Smith}, {O'Brien}, {Sharkey}, {Reddy}, {Springmann}, {Lejoly}, {Volk}, {Conrad}, \& {Veillet}}]{Kareta2019}
{Kareta}, T., {Andrews}, J., {Noonan}, J.~W., {et~al.} 2020, \bibinfo{title}{{Carbon Chain Depletion of 2I/Borisov},} \apjl, 889, L38, \dodoi{10.3847/2041-8213/ab6a08}

\bibitem[{T. {Kareta} {et~al.}(2025){Kareta}, {Champagne}, {McClure}, {Emery}, {Sharkey}, {Bauer}, {Connelley}, {Rayner}, {Thomas}, {Reddy}, \& {Firgard}}]{Kareta2025}
{Kareta}, T., {Champagne}, C., {McClure}, L., {et~al.} 2025, \bibinfo{title}{{Near-discovery Observations of Interstellar Comet 3I/ATLAS with the NASA Infrared Telescope Facility},} \apjl, 990, L65, \dodoi{10.3847/2041-8213/adfbdf}

\bibitem[{L. {Kohoutek}(2001){Kohoutek}}]{2001A&A...378..843K}
{Kohoutek}, L. 2001, \bibinfo{title}{{Version 2000 of the Catalogue of Galactic Planetary Nebulae},} \aap, 378, 843, \dodoi{10.1051/0004-6361:20011162}

\bibitem[{L.~E. {Langland-Shula} \& G.~H. {Smith}(2011){Langland-Shula} \& {Smith}}]{2011Icar..213..280L}
{Langland-Shula}, L.~E., \& {Smith}, G.~H. 2011, \bibinfo{title}{{Comet classification with new methods for gas and dust spectroscopy},} \icarus, 213, 280, \dodoi{10.1016/j.icarus.2011.02.007}

\bibitem[{A.-C. {Levasseur-Regourd} {et~al.}(2018){Levasseur-Regourd}, {Agarwal}, {Cottin}, {Engrand}, {Flynn}, {Fulle}, {Gombosi}, {Langevin}, {Lasue}, {Mannel}, {Merouane}, {Poch}, {Thomas}, \& {Westphal}}]{2018SSRv..214...64L}
{Levasseur-Regourd}, A.-C., {Agarwal}, J., {Cottin}, H., {et~al.} 2018, \bibinfo{title}{{Cometary Dust},} \ssr, 214, 64, \dodoi{10.1007/s11214-018-0496-3}

\bibitem[{W.~G. {Levine} {et~al.}(2021){Levine}, {Cabot}, {Seligman}, \& {Laughlin}}]{Levine2021}
{Levine}, W.~G., {Cabot}, S. H.~C., {Seligman}, D., \& {Laughlin}, G. 2021, \bibinfo{title}{{Constraints on the Occurrence of 'Oumuamua-Like Objects},} \apj, 922, 39, \dodoi{10.3847/1538-4357/ac1fe6}

\bibitem[{W.~G. {Levine} \& G. {Laughlin}(2021){Levine} \& {Laughlin}}]{Levine2021_h2}
{Levine}, W.~G., \& {Laughlin}, G. 2021, \bibinfo{title}{{Assessing the Formation of Solid Hydrogen Objects in Starless Molecular Cloud Cores},} \apj, 912, 3, \dodoi{10.3847/1538-4357/abec85}

\bibitem[{H.~W. {Lin} {et~al.}(2020){Lin}, {Lee}, {Gerdes}, {Adams}, {Becker}, {Napier}, \& {Markwardt}}]{Lin2020}
{Lin}, H.~W., {Lee}, C.-H., {Gerdes}, D.~W., {et~al.} 2020, \bibinfo{title}{{Detection of Diatomic Carbon in 2I/Borisov},} \apjl, 889, L30, \dodoi{10.3847/2041-8213/ab6bd9}

\bibitem[{C.~M. {Lisse} {et~al.}(2025){Lisse}, {Bach}, {Bryan}, {Crill}, {Cukierman}, {Dor{\'e}}, {Fabinsky}, {Faisst}, {Korngut}, {Melnick}, {Rustamkulov}, {Tolls}, {Werner}, {Sitko}, {Champagne}, {Connelley}, {Emery}, {Fernandez}, {Yang}, \& {the SPHEREx Science Team}}]{Lisse2025}
{Lisse}, C.~M., {Bach}, Y.~P., {Bryan}, S., {et~al.} 2025, \bibinfo{title}{{SPHEREx Discovery of Strong Water Ice Absorption and an Extended Carbon Dioxide Coma in 3I/ATLAS},} arXiv e-prints, arXiv:2508.15469, \dodoi{10.48550/arXiv.2508.15469}

\bibitem[{A. {Loeb} {et~al.}(2025){Loeb}, {Hibberd}, \& {Crowl}}]{Loeb2025b}
{Loeb}, A., {Hibberd}, A., \& {Crowl}, A. 2025, \bibinfo{title}{{Intercepting 3I/ATLAS at Closest Approach to Jupiter with the Juno spacecraft},} arXiv e-prints, arXiv:2507.21402.
\newblock \doarXiv{2507.21402}

\bibitem[{J.~X. {Luu} {et~al.}(2020){Luu}, {Flekk{\o}y}, \& {Toussaint}}]{Luu20}
{Luu}, J.~X., {Flekk{\o}y}, E.~G., \& {Toussaint}, R. 2020, \bibinfo{title}{{'Oumuamua as a Cometary Fractal Aggregate: The ``Dust Bunny'' Model},} \apjl, 900, L22, \dodoi{10.3847/2041-8213/abafa7}

\bibitem[{J. {Martinez-Palomera} {et~al.}(2025){Martinez-Palomera}, {Tuson}, {Hedges}, {Dotson}, {Barclay}, \& {Powell}}]{Martinez-Palomera2025}
{Martinez-Palomera}, J., {Tuson}, A., {Hedges}, C., {et~al.} 2025, \bibinfo{title}{{Pre-discovery TESS Observations of Interstellar Object 3I/ATLAS},} arXiv e-prints, arXiv:2508.02499.
\newblock \doarXiv{2508.02499}

\bibitem[{P. {Martini} {et~al.}(2011){Martini}, {Stoll}, {Derwent}, {Zhelem}, {Atwood}, {Gonzalez}, {Mason}, {O'Brien}, {Pappalardo}, {Pogge}, {Ward}, \& {Wong}}]{2011PASP..123..187M}
{Martini}, P., {Stoll}, R., {Derwent}, M.~A., {et~al.} 2011, \bibinfo{title}{{The Ohio State Multi-Object Spectrograph},} \pasp, 123, 187, \dodoi{10.1086/658357}

\bibitem[{A.~J. {McKay} {et~al.}(2020){McKay}, {Cochran}, {Dello Russo}, \& {DiSanti}}]{McKay2020}
{McKay}, A.~J., {Cochran}, A.~L., {Dello Russo}, N., \& {DiSanti}, M.~A. 2020, \bibinfo{title}{{Detection of a Water Tracer in Interstellar Comet 2I/Borisov},} \apjl, 889, L10, \dodoi{10.3847/2041-8213/ab64ed}

\bibitem[{K.~J. {Meech} {et~al.}(2017){Meech}, {Weryk}, {Micheli}, {Kleyna}, {Hainaut}, {Jedicke}, {Wainscoat}, {Chambers}, {Keane}, {Petric}, {Denneau}, {Magnier}, {Berger}, {Huber}, {Flewelling}, {Waters}, {Schunova-Lilly}, \& {Chastel}}]{Meech2017}
{Meech}, K.~J., {Weryk}, R., {Micheli}, M., {et~al.} 2017, \bibinfo{title}{{A brief visit from a red and extremely elongated interstellar asteroid},} Nature, 552, 378, \dodoi{10.1038/nature25020}

\bibitem[{M. {Micheli} {et~al.}(2018){Micheli}, {Farnocchia}, {Meech}, {Buie}, {Hainaut}, {Prialnik}, {Sch{\"o}rghofer}, {Weaver}, {Chodas}, {Kleyna}, {Weryk}, {Wainscoat}, {Ebeling}, {Keane}, {Chambers}, {Koschny}, \& {Petropoulos}}]{Micheli2018}
{Micheli}, M., {Farnocchia}, D., {Meech}, K.~J., {et~al.} 2018, \bibinfo{title}{{Non-gravitational acceleration in the trajectory of 1I/2017 U1 ('Oumuamua)},} Nature, 559, 223, \dodoi{10.1038/s41586-018-0254-4}

\bibitem[{A. Moro-Mart{\'\i}n(2019)Moro-Mart{\'\i}n}]{MoroMartin2019}
Moro-Mart{\'\i}n, A. 2019, \bibinfo{title}{Could 1I/’Oumuamua be an Icy Fractal Aggregate?} \apjl, 872, L32

\bibitem[{A. {Moro-Mart{\'\i}n}(2022){Moro-Mart{\'\i}n}}]{MoroMartin2022}
{Moro-Mart{\'\i}n}, A. 2022, \bibinfo{title}{{Interstellar planetesimals},} arXiv e-prints, arXiv:2205.04277, \dodoi{10.48550/arXiv.2205.04277}

\bibitem[{M.~J. {Mumma} \& S.~B. {Charnley}(2011){Mumma} \& {Charnley}}]{2011ARA&A..49..471M}
{Mumma}, M.~J., \& {Charnley}, S.~B. 2011, \bibinfo{title}{{The Chemical Composition of Comets{\textemdash}Emerging Taxonomies and Natal Heritage},} \araa, 49, 471, \dodoi{10.1146/annurev-astro-081309-130811}

\bibitem[{J.~B. {Oke}(1990){Oke}}]{1990AJ.....99.1621O}
{Oke}, J.~B. 1990, \bibinfo{title}{{Faint Spectrophotometric Standard Stars},} \aj, 99, 1621, \dodoi{10.1086/115444}

\bibitem[{C. {Opitom} {et~al.}(2017){Opitom}, {Snodgrass}, {Fitzsimmons}, {Jehin}, {Manfroid}, {Tozzi}, {Faggi}, \& {Gillon}}]{2017MNRAS.469S.222O}
{Opitom}, C., {Snodgrass}, C., {Fitzsimmons}, A., {et~al.} 2017, \bibinfo{title}{{Ground-based monitoring of comet 67P/Churyumov-Gerasimenko gas activity throughout the Rosetta mission},} \mnras, 469, S222, \dodoi{10.1093/mnras/stx1591}

\bibitem[{C. {Opitom} {et~al.}(2019){Opitom}, {Fitzsimmons}, {Jehin}, {Moulane}, {Hainaut}, {Meech}, {Yang}, {Snodgrass}, {Micheli}, {Keane}, {Benkhaldoun}, \& {Kleyna}}]{Opitom2019}
{Opitom}, C., {Fitzsimmons}, A., {Jehin}, E., {et~al.} 2019, \bibinfo{title}{{2I/Borisov: A C$_{2}$-depleted interstellar comet},} \aap, 631, L8, \dodoi{10.1051/0004-6361/201936959}

\bibitem[{C. {Opitom} {et~al.}(2025){Opitom}, {Snodgrass}, {Jehin}, {Bannister}, {Bufanda}, {Deam}, {Dorsey}, {Ferrais}, {Hmiddouch}, {Knight}, {Kokotanekova}, {Leicester}, {Marsset}, {Murphy}, {Okoth}, {Ridden-Harper}, {Vander Donckt}, {Ferellec}, {Hutsemekers}, {Lippi}, {Manfroid}, \& {Benkhaldoun}}]{Opitom2025}
{Opitom}, C., {Snodgrass}, C., {Jehin}, E., {et~al.} 2025, \bibinfo{title}{{Snapshot of a new interstellar comet: 3I/ATLAS has a red and featureless spectrum},} arXiv e-prints, arXiv:2507.05226, \dodoi{10.48550/arXiv.2507.05226}

\bibitem[{D.~J. {Osip} {et~al.}(2010){Osip}, {A'Hearn}, \& {Raugh}}]{2010PDSS.8133E....O}
{Osip}, D.~J., {A'Hearn}, M., \& {Raugh}, A.~C. 2010, \bibinfo{title}{{Lowell Observatory Cometary Database - Production Rates},}, NASA Planetary Data System, id. EAR-C-PHOT-5-RDR-LOWELL-COMET-DB-PR-V1.0 \dodoi{10.26007/0A3F-R875}

\bibitem[{R.~S. {Park} {et~al.}(2018){Park}, {Pisano}, {Lazio}, {Chodas}, \& {Naidu}}]{Park2018}
{Park}, R.~S., {Pisano}, D.~J., {Lazio}, T. J.~W., {Chodas}, P.~W., \& {Naidu}, S.~P. 2018, \bibinfo{title}{{Search for OH 18 cm Radio Emission from 1I/2017 U1 with the Green Bank Telescope},} \aj, 155, 185, \dodoi{10.3847/1538-3881/aab78d}

\bibitem[{T.~H. {Puzia} {et~al.}(2025){Puzia}, {Rahatgaonkar}, {Carvajal}, {Nayak}, \& {Luco}}]{2025ApJ...990L..27P}
{Puzia}, T.~H., {Rahatgaonkar}, R., {Carvajal}, J.~P., {Nayak}, P.~K., \& {Luco}, B. 2025, \bibinfo{title}{{Spectral Characteristics of Interstellar Object 3I/ATLAS from SOAR Observations},} \apjl, 990, L27, \dodoi{10.3847/2041-8213/adfa0b}

\bibitem[{R. {Rahatgaonkar} {et~al.}(2025){Rahatgaonkar}, {Carvajal}, {Puzia}, {Luco}, {Jehin}, {Hutsem{\'e}kers}, {Opitom}, {Manfroid}, {Marsset}, {Yang}, {Buchanan}, {Fraser}, {Forbes}, {Bannister}, {Bodewits}, {Bolin}, {Belyakov}, {Knight}, {Snodgrass}, {Bufanda}, {Dorsey}, {Ferellec}, {La Forgia}, {Lippi}, {Murphy}, {Nayak}, \& {Vander Donckt}}]{Rahatgaonkar2025}
{Rahatgaonkar}, R., {Carvajal}, J.~P., {Puzia}, T.~H., {et~al.} 2025, \bibinfo{title}{{VLT observations of interstellar comet 3I/ATLAS II. From quiescence to glow: Dramatic rise of Ni I emission and incipient CN outgassing at large heliocentric distances},} arXiv e-prints, arXiv:2508.18382, \dodoi{10.48550/arXiv.2508.18382}

\bibitem[{J.~P. Sanchez \& C. Snodgrass(2025)Sanchez \& Snodgrass}]{Sanchez_2025}
Sanchez, J.~P., \& Snodgrass, C. 2025, \bibinfo{title}{Analysis of Trajectories to 3I/ATLAS with a Comet Interceptor-like Spacecraft,} Research Notes of the AAS, 9, 207, \dodoi{10.3847/2515-5172/adf4c4}

\bibitem[{T. {Santana-Ros} {et~al.}(2025){Santana-Ros}, {Ivanova}, {Mykhailova}, {Erasmus}, {Kami{\'n}ski}, {Oszkiewicz}, {Kwiatkowski}, {Hus{\'a}rik}, {Ngwane}, \& {Penttil{\"a}}}]{2025arXiv250800808S}
{Santana-Ros}, T., {Ivanova}, O., {Mykhailova}, S., {et~al.} 2025, \bibinfo{title}{{Temporal Evolution of the Third Interstellar Comet 3I/ATLAS: Spin, Color, Spectra and Dust Activity},} arXiv e-prints, arXiv:2508.00808, \dodoi{10.48550/arXiv.2508.00808}

\bibitem[{D. {Schleicher}(2025){Schleicher}}]{2025ATel17352}
{Schleicher}, D. 2025, \bibinfo{title}{{The Detection of CN in Interstellar Comet 3I/ATLAS},} The Astronomer's Telegram, 17352

\bibitem[{D. {Schleicher} \& A. {Bair}(2014){Schleicher} \& {Bair}}]{2014acm..conf..475S}
{Schleicher}, D., \& {Bair}, A. 2014, in Asteroids, Comets, Meteors 2014, 475

\bibitem[{D.~G. {Schleicher}(2010){Schleicher}}]{2010AJ....140..973S}
{Schleicher}, D.~G. 2010, \bibinfo{title}{{The Fluorescence Efficiencies of the CN Violet Bands in Comets},} \aj, 140, 973, \dodoi{10.1088/0004-6256/140/4/973}

\bibitem[{D.~G. {Schleicher} \& A.~N. {Bair}(2011){Schleicher} \& {Bair}}]{2011AJ....141..177S}
{Schleicher}, D.~G., \& {Bair}, A.~N. 2011, \bibinfo{title}{{The Composition of the Interior of Comet 73P/Schwassmann-Wachmann 3: Results from Narrowband Photometry of Multiple Components},} \aj, 141, 177, \dodoi{10.1088/0004-6256/141/6/177}

\bibitem[{Z. {Sekanina}(2019){Sekanina}}]{Sekanina2019}
{Sekanina}, Z. 2019, \bibinfo{title}{{Outgassing As Trigger of 1I/`Oumuamua's Nongravitational Acceleration: Could This Hypothesis Work at All?},} arXiv e-prints, arXiv:1905.00935.
\newblock \doarXiv{1905.00935}

\bibitem[{D. {Seligman} \& G. {Laughlin}(2020){Seligman} \& {Laughlin}}]{Seligman2020}
{Seligman}, D., \& {Laughlin}, G. 2020, \bibinfo{title}{{Evidence that 1I/2017 U1 ('Oumuamua) was Composed of Molecular Hydrogen Ice},} \apjl, 896, L8, \dodoi{10.3847/2041-8213/ab963f}

\bibitem[{D.~Z. {Seligman} \& A. {Moro-Mart{\'\i}n}(2023){Seligman} \& {Moro-Mart{\'\i}n}}]{Seligman2023}
{Seligman}, D.~Z., \& {Moro-Mart{\'\i}n}, A. 2023, \bibinfo{title}{{Interstellar objects},} Contemporary Physics, 63, 200, \dodoi{10.1080/00107514.2023.2203976}

\bibitem[{D.~Z. {Seligman} {et~al.}(2022){Seligman}, {Rogers}, {Cabot}, {Noonan}, {Kareta}, {Mandt}, {Ciesla}, {McKay}, {Feinstein}, {Levine}, {Bean}, {Nordlander}, {Krumholz}, {Mansfield}, {Hoover}, \& {Van Clepper}}]{2022PSJ.....3..150S}
{Seligman}, D.~Z., {Rogers}, L.~A., {Cabot}, S. H.~C., {et~al.} 2022, \bibinfo{title}{{The Volatile Carbon-to-oxygen Ratio as a Tracer for the Formation Locations of Interstellar Comets},} \psj, 3, 150, \dodoi{10.3847/PSJ/ac75b5}

\bibitem[{D.~Z. {Seligman} {et~al.}(2023){Seligman}, {Farnocchia}, {Micheli}, {Vokrouhlick{\'y}}, {Taylor}, {Chesley}, {Bergner}, {Vere{\v{s}}}, {Hainaut}, {Meech}, {Devogele}, {Pravec}, {Matson}, {Deen}, {Tholen}, {Weryk}, {Rivera-Valent{\'\i}n}, \& {Sharkey}}]{Seligman2023b}
{Seligman}, D.~Z., {Farnocchia}, D., {Micheli}, M., {et~al.} 2023, \bibinfo{title}{{Dark Comets? Unexpectedly Large Nongravitational Accelerations on a Sample of Small Asteroids},} \psj, 4, 35, \dodoi{10.3847/PSJ/acb697}

\bibitem[{D.~Z. {Seligman} {et~al.}(2024){Seligman}, {Farnocchia}, {Micheli}, {Hainaut}, {Hsieh}, {Feinstein}, {Chesley}, {Taylor}, {Masiero}, \& {Meech}}]{Seligman2024PNAS}
{Seligman}, D.~Z., {Farnocchia}, D., {Micheli}, M., {et~al.} 2024, \bibinfo{title}{{Two distinct populations of dark comets delineated by orbits and sizes},} Proceedings of the National Academy of Science, 121, e2406424121, \dodoi{10.1073/pnas.2406424121}

\bibitem[{D.~Z. {Seligman} {et~al.}(2025){Seligman}, {Micheli}, {Farnocchia}, {Denneau}, {Noonan}, {Hsieh}, {Santana-Ros}, {Tonry}, {Auchettl}, {Conversi}, {Devog{\`e}le}, {Faggioli}, {Feinstein}, {Fenucci}, {Ferrais}, {Frincke}, {Gillon}, {Hainaut}, {Hart}, {Hoffman}, {Holt}, {Hoogendam}, {Huber}, {Jehin}, {Kareta}, {Keane}, {Kelley}, {Lister}, {Mandt}, {Manfroid}, {Mar{\v{c}}eta}, {Meech}, {Amine Miftah}, {Morgan}, {Oca{\~n}a}, {Pe{\~n}a-Asensio}, {Shappee}, {Siverd}, {Taylor}, {Tucker}, {Wainscoat}, {Weryk}, {Wray}, {Yaginuma}, {Yang}, {Ye}, \& {Zhang}}]{Seligman2025}
{Seligman}, D.~Z., {Micheli}, M., {Farnocchia}, D., {et~al.} 2025, \bibinfo{title}{{Discovery and Preliminary Characterization of a Third Interstellar Object: 3I/ATLAS},} \apjl, 989, L36, \dodoi{10.3847/2041-8213/adf49a}

\bibitem[{C. {Snodgrass} \& G.~H. {Jones}(2019){Snodgrass} \& {Jones}}]{Snodgrass2019}
{Snodgrass}, C., \& {Jones}, G.~H. 2019, \bibinfo{title}{{The European Space Agency's Comet Interceptor lies in wait},} Nature Communications, 10, 5418, \dodoi{10.1038/s41467-019-13470-1}

\bibitem[{R. {Stoll} {et~al.}(2010){Stoll}, {Martini}, {Derwent}, {Gonzalez}, {O'Brien}, {Pappalardo}, {Pogge}, {Wong}, \& {Zhelem}}]{2010SPIE.7735E..4LS}
{Stoll}, R., {Martini}, P., {Derwent}, M.~A., {et~al.} 2010, in Society of Photo-Optical Instrumentation Engineers (SPIE) Conference Series, Vol. 7735, Ground-based and Airborne Instrumentation for Astronomy III, ed. I.~S. {McLean}, S.~K. {Ramsay}, \& H.~{Takami}, 77354L, \dodoi{10.1117/12.857893}

\bibitem[{A.~G. Taylor \& D.~Z. Seligman(2025)Taylor \& Seligman}]{Taylor2025a}
Taylor, A.~G., \& Seligman, D.~Z. 2025, \bibinfo{title}{The {Kinematic} {Age} of {3I}/{ATLAS} and its {Implications} for {Early} {Planet} {Formation},} arXiv, \dodoi{10.48550/arXiv.2507.08111}

\bibitem[{J.~L. {Tonry} {et~al.}(2018){Tonry}, {Denneau}, {Heinze}, {Stalder}, {Smith}, {Smartt}, {Stubbs}, {Weiland}, \& {Rest}}]{Tonry2018a}
{Tonry}, J.~L., {Denneau}, L., {Heinze}, A.~N., {et~al.} 2018, \bibinfo{title}{{ATLAS: A High-cadence All-sky Survey System},} \pasp, 130, 064505, \dodoi{10.1088/1538-3873/aabadf}

\bibitem[{D.~E. {Trilling} {et~al.}(2018){Trilling}, {Mommert}, {Hora}, {Farnocchia}, {Chodas}, {Giorgini}, {Smith}, {Carey}, {Lisse}, {Werner}, {McNeill}, {Chesley}, {Emery}, {Fazio}, {Fernandez}, {Harris}, {Marengo}, {Mueller}, {Roegge}, {Smith}, {Weaver}, {Meech}, \& {Micheli}}]{Trilling2018}
{Trilling}, D.~E., {Mommert}, M., {Hora}, J.~L., {et~al.} 2018, \bibinfo{title}{{Spitzer Observations of Interstellar Object 1I/{\textquoteleft}Oumuamua},} \aj, 156, 261, \dodoi{10.3847/1538-3881/aae88f}

\bibitem[{M. {Weiler}(2012){Weiler}}]{2012A&A...538A.149W}
{Weiler}, M. 2012, \bibinfo{title}{{The chemistry of C$_{3}$ and C$_{2}$ in cometary comae. I. Current models revisited},} \aap, 538, A149, \dodoi{10.1051/0004-6361/201117480}

\bibitem[{F.~L. {Whipple}(1950){Whipple}}]{1950ApJ...111..375W}
{Whipple}, F.~L. 1950, \bibinfo{title}{{A comet model. I. The acceleration of Comet Encke},} \apj, 111, 375, \dodoi{10.1086/145272}

\bibitem[{G.~V. {Williams} {et~al.}(2017){Williams}, {Sato}, {Sarneczky}, {Wainscoat}, {Woodworth}, \& {Meech}}]{Williams17}
{Williams}, G.~V., {Sato}, H., {Sarneczky}, K., {et~al.} 2017, \bibinfo{title}{{Minor Planets 2017 SN\_33 and 2017 U1},} Central Bureau Electronic Telegrams, 4450, 1

\bibitem[{C. {Wolf} {et~al.}(2018){Wolf}, {Onken}, {Luvaul}, {Schmidt}, {Bessell}, {Chang}, {Da Costa}, {Mackey}, {Martin-Jones}, {Murphy}, {Preston}, {Scalzo}, {Shao}, {Smillie}, {Tisserand}, {White}, \& {Yuan}}]{skymapper_dr1}
{Wolf}, C., {Onken}, C.~A., {Luvaul}, L.~C., {et~al.} 2018, \bibinfo{title}{{SkyMapper Southern Survey: First Data Release (DR1)},} \pasa, 35, e010, \dodoi{10.1017/pasa.2018.5}

\bibitem[{Z. {Xing} {et~al.}(2020){Xing}, {Bodewits}, {Noonan}, \& {Bannister}}]{Xing2020}
{Xing}, Z., {Bodewits}, D., {Noonan}, J., \& {Bannister}, M.~T. 2020, \bibinfo{title}{{Water Production Rates and Activity of Interstellar Comet 2I/Borisov},} \apjl, 893, L48, \dodoi{10.3847/2041-8213/ab86be}

\bibitem[{Z. {Xing} {et~al.}(2025){Xing}, {Oset}, {Noonan}, \& {Bodewits}}]{Xing2025}
{Xing}, Z., {Oset}, S., {Noonan}, J., \& {Bodewits}, D. 2025, \bibinfo{title}{{Water Production Rates of the Interstellar Object 3I/ATLAS},} \apjl, 991, L50, \dodoi{10.3847/2041-8213/ae08ab}

\bibitem[{A. {Yaginuma} {et~al.}(2025){Yaginuma}, {Frincke}, {Seligman}, {Mandt}, {DellaGiustina}, {Pe{\~n}a-Asensio}, {Taylor}, \& {Nolan}}]{Yaginuma2025}
{Yaginuma}, A., {Frincke}, T., {Seligman}, D.~Z., {et~al.} 2025, \bibinfo{title}{{The Feasibility of a Spacecraft Flyby with the Third Interstellar Object 3I/ATLAS from Earth or Mars},} arXiv e-prints, arXiv:2507.15755, \dodoi{10.48550/arXiv.2507.15755}

\bibitem[{B. Yang {et~al.}(2025)Yang, Meech, Connelley, Zhao, \& Keane}]{Yang2025}
Yang, B., Meech, K.~J., Connelley, M., Zhao, R., \& Keane, J.~V. 2025, \bibinfo{title}{Spectroscopic Characterization of Interstellar Object 3I/ATLAS: Water Ice in the Coma,} The Astrophysical Journal Letters, 992, L9, \dodoi{10.3847/2041-8213/ae08a7}

\bibitem[{B. Yang {et~al.}(2021)Yang, Li, Cordiner, Chang, Hainaut, Williams, Meech, Keane, \& Villard}]{yang2021}
Yang, B., Li, A., Cordiner, M.~A., {et~al.} 2021, \bibinfo{title}{Compact pebbles and the evolution of volatiles in the interstellar comet 2I/Borisov,} Nature Astronomy, \dodoi{10.1038/s41550-021-01336-w}

\bibitem[{Q.-Z. {Ye} {et~al.}(2017){Ye}, {Zhang}, {Kelley}, \& {Brown}}]{Ye2017}
{Ye}, Q.-Z., {Zhang}, Q., {Kelley}, M.~S.~P., \& {Brown}, P.~G. 2017, \bibinfo{title}{{1I/2017 U1 (`Oumuamua) is Hot: Imaging, Spectroscopy, and Search of Meteor Activity},} Astrophysical Journal Letters, 851, L5, \dodoi{10.3847/2041-8213/aa9a34}

\end{thebibliography}
\bibliographystyle{aasjournal}



\end{document}